\documentclass[a4paper,fleqn,usenatbib]{mnras}
\usepackage{newtxtext,newtxmath}
\usepackage[T1]{fontenc}
\usepackage{ae,aecompl}

\usepackage{graphicx}	
\usepackage{amsmath}	
\usepackage{amssymb}	

\usepackage{float}
\usepackage{units}
\usepackage{siunitx} 
\usepackage[version=4]{mhchem}

\usepackage{tikz,pgfplots}
       \usetikzlibrary{arrows,arrows.meta}
       \usetikzlibrary{decorations.pathreplacing}
      \usetikzlibrary{pgfplots.groupplots}
       \usetikzlibrary{spy}
       
\usepackage{accents}
\usepackage{xfrac}
\pgfplotsset{compat=1.14}

\usepackage{color}
\definecolor{mycolor1}{RGB}{ 0,113.9850,188.9550}
\definecolor{mycolor2}{RGB}{216.7500,82.8750,24.9900} 
\definecolor{mycolor3}{RGB}{236.8950,7,3.8750} 
\definecolor{mycolor4}{RGB}{125.9700,46.9200,141.7800} 
\definecolor{mycolor5}{RGB}{118.8300,171.8700,47.9400} 
\definecolor{mygreen}{RGB}{212,187,106} 
\definecolor{mylilas}{RGB}{18,54,82}
\definecolor{matlabmygreen}{RGB}{28,172,0} 
\definecolor{matlabmylilas}{RGB}{170,55,241}
\definecolor{mycolor1}{rgb}{1.0,0.729,0.369}
\definecolor{mycolor2}{rgb}{0.122,0.235,0.325}

\title[Solar activity simulation and forecast]{Solar activity simulation and forecast with a flux-transport dynamo}

\author[A.Macario-Rojas, K.L.Smith, and P.C.E.Roberts]{
Alejandro Macario-Rojas,\thanks{E-mail: alejandromacario.rojas@manchester.ac.uk}
Katharine L. Smith,
Peter C. E. Roberts
\\
School of Mechanical Aerospace and Civil Engineering, The University of Manchester, M13 9PL, UK\\
}

\date{Accepted XXX. Received YYY; in original form ZZZ}

\pubyear{2018}

\begin{document}
\label{firstpage}
\pagerange{\pageref{firstpage}--\pageref{lastpage}}
\maketitle

\begin{abstract}

We present the assessment of a diffusion-dominated mean field axisymmetric dynamo model in reproducing historical solar activity and forecast for solar cycle 25. Previous studies point to the Sun's polar magnetic field as an important proxy for solar activity prediction. Extended research using this proxy has been impeded by reduced observational data record only available from 1976. However, there is a recognised need for a solar dynamo model with ample verification over various activity scenarios to improve theoretical standards. The present study aims to explore the use of helioseismology data and reconstructed solar polar magnetic field, to foster the development of robust solar activity forecasts. The research is based on observationally inferred differential rotation morphology, as well as observed and reconstructed polar field using artificial neural network methods via the hemispheric sunspot areas record. Results show consistent reproduction of historical solar activity trends with enhanced results by introducing a precursor rise time coefficient. A weak solar cycle 25, with slow rise time and maximum activity $-14.4 \%$ ($\pm 19.5\%$) with respect to the current cycle 24 is predicted.

\end{abstract}


\section{Introduction} \label{sec:intro}

The application of theoretical dynamo models that aim to reproduce aspects of solar activity is based upon the current understanding of observed and inferred solar phenomena. Despite the significant efforts to produce self-consistent models that can faithfully represent solar activity, the highly complex underlying mechanisms and voids in solar physics knowledge have to date prevented theorists from doing so.  Current wisdom points to the application of varied nonlinear three-dimensional processes converging on observed phenomena, whose large-scale overall evolution is remarkably systematic; this allows, under specific circumstances discussed below, the axisymmetrisation of leading phenomena simplifying the modelling problem.

Axisymmetric solar dynamo models used for forecasts are related by magnetohydrodynamic (MHD) theory, the main difference between these models are the parameterisations that result in differing dynamo characteristics. In general terms, in order to reproduce the main temporo-spatial characteristics of the observed Sun, dynamo model calibration is necessary. The identification of plausible combinations of parameters yielding acceptable reproduction of observed activity is an arduous endeavour that does not necessarily guarantee forecast reliability. The main limitation in this respect, is the narrow time record of relevant solar activity features that can provide suitable input parameters for MHD models in addition to validation data for the outputs of these models. For instance, the approach presented by \citet{jiang2007solar} using observed photospheric magnetic field data as the main parameter for prediction, correctly anticipated the magnitude of the solar cycle 24 and satisfactorily reproduced the three precedent cycles, that is for the cycles where photospheric magnetic field data was available. Some attempts to overcome limitations of solar activity data use other proxies such as the relatively large sunspot number (SSN) record to calibrate solar dynamo models. In this regard, \citet{dikpati2008predicting} and~\citet{karak2010importance} used SSN cycle periods to adjust the dynamo model meridional flow speed, and~\citet{kitiashvili2008application} fitted simplified dynamo equations to observed SSN activity.

The solar activity forecast approach presented in this investigation is centred on reproducing mean large-scale solar activity features using a dynamo model. The preferred solar activity leading indicator is the solar poloidal magnetic field strength in a similar approach to \citet{jiang2007solar}. However, in this investigation we propose the use of polar magnetic field, and the selection of suitable dynamo parameterisations based on the application of a detailed axial solar rotation profile inferred from helioseismology. Additionally, in order to enrich the typical dynamo calibration process, the short record of solar polar magnetic field strength is extended by means of neural network reconstruction. The investigation is organised into two main sections. The following section \ref{sec:theorymhd} describes the mean field axisymmetric dynamo model and its parameterisations, as well as characteristic model assumptions and caveats. Section \ref{sec:solarcyfore} presents the characteristics of the resulting solar-like dynamo, the proposed solar polar field strength reconstruction, and finally the performance of the overall approach in reproducing solar activity is discussed.

\section{The Mean Field Axisymmetric Dynamo Model} \label{sec:theorymhd}

A key characteristic found in mean field axisymmetric dynamo models is the solution of the dynamo problem equations in a meridional slab with suitable boundary conditions. In this investigation a finite differences computational code is developed incorporating the openly-available Fortran SURYA code~\citep{choudhuri2005user} to solve the system of coupled differential equations. The MATLAB\textsuperscript{\tiny\textregistered} code used in this investigation employs separated poloidal and toroidal magnetic diffusivity models as in the SURYA code, and by extension by \citet{jiang2007solar}, as well as the same approach for magnetic growth suppression. However, the solar model parameterisations approach is distinct in this investigation. Since the solar plasma's angular velocity differential rotation profile is well constrained by helioseismology, a realistic parameterisation model is chosen over the typical analytical fit. This in turn underpins the adjustment of the loosely defined solar magnetic diffusivity and poloidal field source parameterisation models enabling the successful reproduction of major solar activity features. The impact of the proposed modifications of the parametrisations on the reproduction of solar activity are discussed after a detailed presentation of the models. In first instance the bulk field dynamics are discussed in subsections \ref{sec:diffrotmodel} and \ref{sec:meridcirc}; subsequently the polar field source and magnetic diffusivity models are presented in subsection \ref{ssec:ftdt}.

\subsection{Angular Velocity} \label{sec:diffrotmodel}

The dominant angular velocity, hereinafter referred to as $\Omega$, and weaker meridional circulation, define the bulk velocity field vector ($\pmb{v}_T$) used in mean field axisymmetric dynamo models. Eq.~\eqref{plasmveltot} summarises the velocity field vector in spherical coordinates in terms of the axial rotation (azimuthal) and the meridional circulation component $\pmb{v}$, where $r$ and $\theta$ are the local radius and colatitude respectively.
\begin{equation}\label{plasmveltot}
\pmb{v}_T  = r \sin \theta~\Omega\left(r,\theta\right)\pmb{\hat{e}}_{\phi} + \pmb{v} 
\end{equation}
The main enigma in solar dynamo theory is how the magnetic field is sustained against diffusion and recycled periodically. A natural explanation for the generation of the azimuthal (toroidal) field from the meridional (poloidal) field is that axial rotation stretches out the local polar field lines frozen to the moving plasma; the resulting wound magnetic field builds the toroidal field ($\omega$-effect). Thus, regions of large shear concentrate and strengthen the toroidal field.

\begin{figure}
  \begin{center}
   \includegraphics[height=5in]{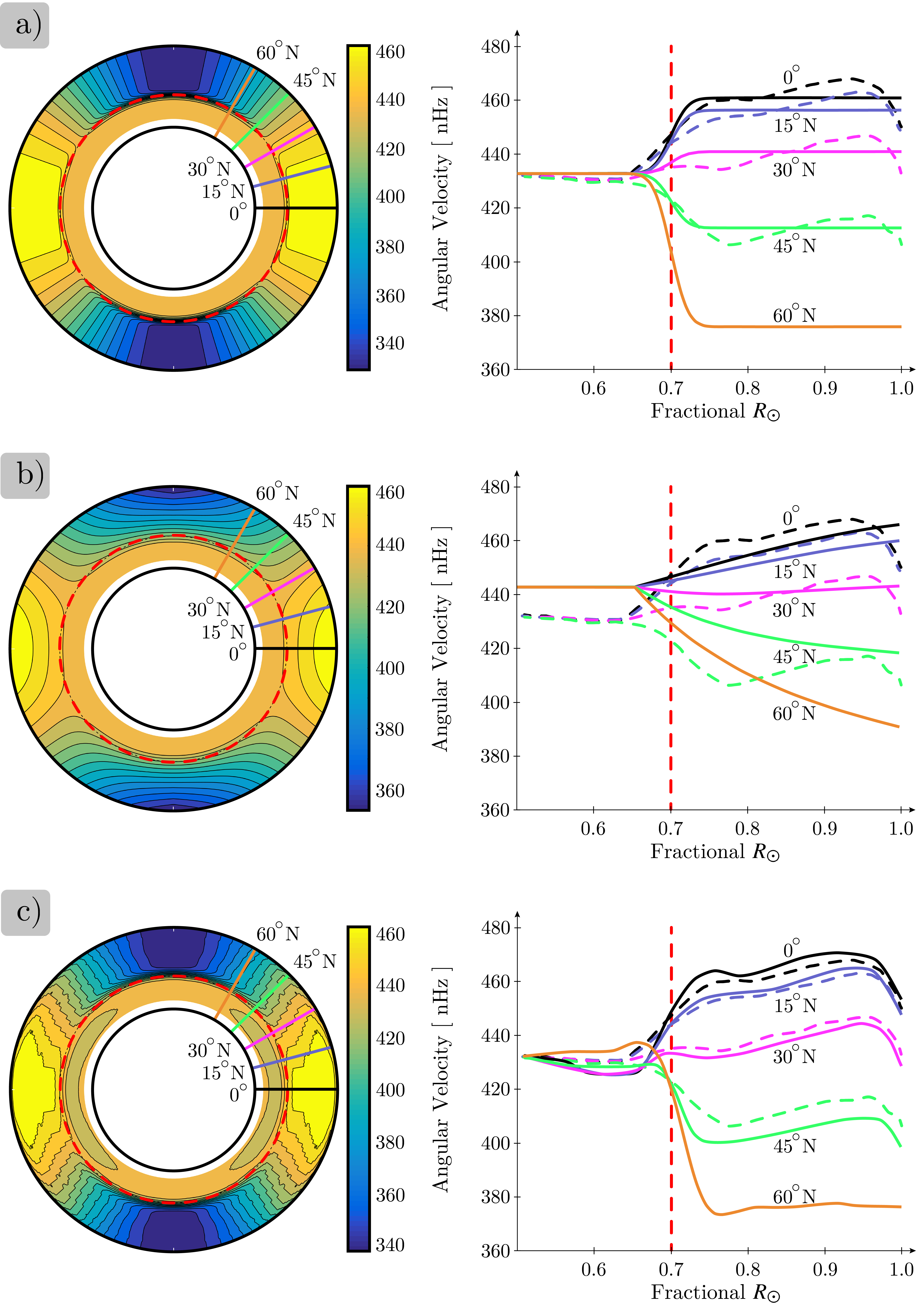}
   \caption{Solar longitudinal averaged isorotation contours (left column) and depth profiles (right column) for selected solar latitudes. The subfigures show examples of axial solar rotation models: a) Analytical fit commonly used in dynamo modelling, b) Mathematical model based on the thermal wind equation, and c) Mapping from helioseismology (GONG). Vertical lines in the depth profiles show the core location of the tachocline, solid curves are the profiles from the models, and dashed curves correspond to inferred observational data.}
    \label{fig:diffrotall}
  \end{center}
\end{figure}

Figure~\ref{fig:diffrotall} shows three representations of the longitudinal averaged rotation map as function of solar latitude and depth. The most common rotation map \citep{charbonneau1999helioseismic} used in solar dynamo modelling is shown in the analytical envelope in Figure~\ref{fig:diffrotall}a. In this subfigure, the comparison of depth profiles for various latitudes between the analytical fit model and inferred data via helioseismology, highlights a rough angular velocity approximation. On the other hand, the profile in Figure~\ref{fig:diffrotall}b is derived from insightful mathematical laws~\citep{balbus2012global}, wherein differential rotation is explained through angular entropy gradients and inertial angular velocity. Finally, Figure~\ref{fig:diffrotall}c shows an O-grid mesh fitted to data from the Global Oscillation Network Group (GONG)~\citep{munoz2009helioseismic}. This realistic model captures most of the characteristics of the average differential rotation morphology which, according to the isorotation contour patterns in Figure~\ref{fig:diffrotall}c, can be divided into three main concentric rings.

Starting from $0.5 R_{\sun}$ towards the surface, an innermost ring shows near constant angular velocity, i.e. solid body rotation, which is abruptly wrenched at $\sim 0.7 R_{\sun}$ (the tachocline) with various magnitudes, depending on latitude. A mid ring is bounded by the tachocline as inner radius up to $\sim 0.95 R_{\sun}$. Within this ring isorotation contours diverge from radial spokes as latitudes approach the solar equator, this is in contrast to the observed trend in Figure~\ref{fig:diffrotall}a, where no divergence with latitude is observed. In the outer ring extending from $0.95 R_{\sun}$ to the surface, the isorotation contours shown in Figure~\ref{fig:diffrotall}c gradually match the observed surface rotation profile; it is suspected that this outer ring is dominated by combinations of vigorous turbulent stresses and magnetic braking~\citep{balbus2012global} reducing angular rotation speed. Due to the fundamental importance of the plasma flow morphology to the current understanding of the solar dynamo, this investigation departs from the typical analytical model in Figure~\ref{fig:diffrotall}a and employs the model shown in Figure~\ref{fig:diffrotall}c. Implications of the application of this differential rotation model for other dynamo parameterisations are reported accordingly in the following subsections.

\subsection{Meridional Circulation}\label{sec:meridcirc}

Vigorous rotation acting on anisotropic and compressible plasma in the convection zone generates a second important macroscopic flow. It is believed that the combined effect of plasma in differential axial rotation, pressure gradients, and turbulent buoyant convective plasma generate cellular circulation that collectively may be the cause of the observed meridional circulation on the solar surface. Analysis of the effect of multiple cellular circulation zones on the reproduction of observed solar activity is studied by~\citet{hazra2014deep}. They concluded that as long as equatorward flow exists at the bottom of the convection zone at low solar latitudes, where azimuthal magnetic field is possibly generated, solar activity can be reproduced effectively in solar dynamo models. In addition, \citet{hazra2014deep} assessed the effect of meridional flow speed variation on the solar cycle period in an array of radially stacked cells. The study indicates a significant influence of the innermost cell speed in determining the length of the solar cycle.

Observed solar surface meridional circulation indicates that this may vary with the progress of the solar cycle, e.g. at mid latitudes $\sim \SI{10}{\metre\per\second}$ during the $23^{rd}$ solar maximum and $\sim \SI{14}{\metre\per\second}$ by the end of that cycle~\citep{hathaway2014solar}. In this respect, the \citet{karak2010importance} study suggests that meridional flow fluctuation may be a fundamental phenomenon influencing the duration and magnitude of solar cycles, or at least an indicator of inner activity. The current definition of reduced meridional flow parameters from observational data, makes the use of the stable single hemispheric cell approach, which is the customary choice in coarse reproduction of solar-like activity. In this investigation we adopt this standard approach acknowledging that this would imply some inherent degree of uncertainty in solar activity reproduction and forecasting. However this is justified by the fact that a single hemispheric cell encompasses the essential meridional flow component of the most up to date representation of the solar dynamo.

The two-dimensional meridional circulation with velocity $\pmb{v}$ can be expressed through a stream function $\psi$ as shown in equation~\eqref{denvel} assuming polytropic density stratification given by equation~\eqref{denstrat}. In this regard, \citet{cardoso2012impact} show that a more realistic description of density stratification may lead to noticeable changes in plasma flow evolution. However, considering the existing uncertainty produced by the meridional circulation structure, the standard polytropic stratification is deemed sufficient for the axisymmetric dynamo model used in this investigation. 
\begin{equation}\label{denvel}
\rho \pmb{v}  = \pmb{\nabla} \pmb{\times} \left[\psi \left(r,\theta\right)\pmb{\hat{e}}_{\phi}\right]
\end{equation}
\begin{equation}\label{denstrat}
\rho = \left(\frac{R_{\sun}}{r} - \gamma \right)^m
\end{equation}
\begin{equation}  \label{streamf}
\begin{split}
\psi r \sin \theta  = & ~\psi_0 \left( r - R_p \right) \sin \left[ \frac{\pi \left( r - R_p \right)}{\left( R_{\sun} - R_p \right)} \right] \left[ 1 - e^{-\beta_1 \theta^\epsilon}\right]  \\
& \times\left[1 - e^{\beta_2  \left(\theta - \sfrac{\pi}{2} \right)}\right] e^{\left[\sfrac{\left(r - r_0 \right)}{\Gamma}\right]^2}
\end{split}
\end{equation}
\begin{table}
\renewcommand{\thetable}{\arabic{table}}
\caption{Meridional circulation parameters.} \label{table:T_meridpar}
\begin{tabular}{lccc}
\hline
Parameter & Reference model & This model & Unit\\
\hline
$\beta_1$            &      $1.36 \times 10^{-8}$  & $1.50 \times 10^{-8}$  &   $\SI{}{\per\metre}$\\
$\beta_2$            &      $1.63 \times 10^{-8}$  &$1.80 \times 10^{-8}$  &  $\SI{}{\per\metre}$\\   
$\gamma$           &       0.95      & 0.95      &       $-$          \\
$\Gamma$          &       $3.47\times 10^8$  &$3.47\times 10^8$  & $\SI{}{\metre}$\\
$\epsilon$           &   2.0000001  &2.0000001  &        $-$         \\
$m$                     &          1.5       &1.5       &        $-$       \\
$R_{p}$              & $0.61 R_{\sun}$ & $0.61 R_{\sun}$ &    $\SI{}{\metre}$          \\
$v_{0}$              & $-29$$^a$ & $-25$ &    $\SI{}{\metre\per\second}$          \\
\hline
\multicolumn{3}{l}{$^a$ Errata reported in \citet{jiang2007solar}}\\
\end{tabular}
\end{table}
The hemispheric stream function model in equation~\eqref{streamf}~\citep{chatterjee2004full} employs the parameters reported in Table~\ref{table:T_meridpar}. With the exception of $\beta$ values and poleward flow velocity \textit{near} the surface at mid-latitudes, $v_{0}$, the parameters used in this investigation are equal to those reported by~\citet{chatterjee2004full}, hereinafter referred to as the reference model. It is important to bear in mind that variation in parameter values between this investigation and the reference model result from the use of different differential rotation maps as discussed in subsection \ref{sec:diffrotmodel}. As the reproduction of solar-like activity depends on the characteristics of the merged parameterisations, similar variations are expected to result for all parameterisations of the solar dynamo model.

The meridional circulation model used in this investigation generates a maximum equatorward counterflow of $\SI{2.29}{\metre\per\second}$ at the tachocline region, which is consistent with the observed trend in equatorward sunspot drift near midlatitudes~\citep{hathaway2003evidence}. In addition, the model yields a reference Schwabe cycle of $10.87$ years in agreement to the average cycle of interest in this investigation and similar to that reported by ~\citet{jiang2007solar} ($10.80$ years). 

\subsection{Poloidal Field Source and Magnetic Diffusivity Models} \label{ssec:ftdt}

The generation of the poloidal magnetic field is of significance to the current theoretical explanation of the solar dynamo. The origin of this magnetic field component can be ascribed to small-scale toroidal flux tube rotation under the action of Coriolis forces (decay of active regions by cyclonic turbulence termed the $\alpha$-effect), the body-axial flux tube twist contributing to the poloidal component~\citep{priest2014magnetohydrodynamics}, possibly by magnetic pumping at the base of the convection zone, and predominantly by diffusion and reconnection of near-surface tilted dipolar regions viz. the Babcock-Leighton process. These chiefly near-surface phenomena are argued to be the source of the poloidal field regeneration. The stochastic nature of the poloidal field generation dominates the part of the Schwabe cycle, from solar maximum to the end of the cycle, and is believed to be one of the most important causes of solar cycle irregularities~\citep{jiang2007solar}. This theory is supported by the observed regularity and predictability of the rising phase of SSN cycles vis-\`{a}-vis the declining (Babcock-Leighton dominated) unstable phase. In continuation with the magnetic cycle, the generated poloidal magnetic field is transported to the poles and solar interior by meridional flow and magnetic diffusion around evolving photospheric convection cells (supergranule cells).

Another identified implication derived from the $\Omega$ model selection and subsequent cumulative modifications, is the required explicit suppression of the poloidal field at the dynamo poles. The Babcock-Leighton process in this investigation is phenomenologically represented by $\alpha$ in equation~\eqref{alphaprof}, modified\footnote{In \citet{hotta2010solar}, the argument of the exponential function in the poloidal source equation includes the term $-\pi/4$.} from~\citet{hotta2010solar}. In this equation the generation of the poloidal field is reasonably concentrated above $0.95 R_{\sun}$, and suppressed at the poles. This $\alpha$ model differs from that used in the reference model, and the SURYA code, where only the latitudinal dependence of the Coriolis effect (represented by the factor $\cos \theta$ in equation~\eqref{alphaprof}) is considered. The suppression of $\alpha$ at the poles and the heuristic value of $\alpha_0  =  \SI{15}{\metre\per\second}$ ensures the relaxation of the dynamo model in this investigation to a stable periodic solution from an initial arbitrary state (super-critical $\alpha$). 
\begin{equation} \label{alphaprof}
\begin{split}
\alpha  =& ~\frac{\alpha_0}{4} \cos \theta \left[1 + \mathrm{erf}\, \left( \frac{r-0.95 R_{\sun}}{0.05 R_{\sun}}  \right) \right]\left[1 - \mathrm{erf}\, \left( \frac{r - R_{\sun}}{0.01 R_{\sun}}  \right) \right]\\
&\times \underbrace{\left(\frac{1}{1+ e^{-30\theta}}\right) \sin \theta} _{\text{Suppression at poles}}
\end{split}
\end{equation}
The final fundamental ingredient in the reproduction of macroscopic solar fluctuations, is the turbulent magnetic diffusivity in the convection zone. The importance of the solar turbulent diffusivity resides in the interconnecting role between the polar and toroidal fields, which are believed to be generated in separate specific regions of the Sun. Unfortunately, the solar turbulent diffusivity process is currently poorly understood and therefore poorly constrained in solar dynamo models. This uncertainty has lead to two main schools of thought considering the diffusivity behaviour in the Sun's interior and the representation of this in mean-field dynamo models. With the exception of its near-surface value, where there is apparent consensus supported by mixing-length theory, the adoption of either high or low turbulent magnetic diffusivity values in solar dynamo models is used~\citet{munoz2010magnetic}. Because there is little evidence of the phenomena taking place in the Sun's interior, flux transport dynamo parameters are adjusted within characteristic ranges to match surface phenomena that in both diffusivity cases may reproduce observed patterns. However, the implications for solar activity forecasting resulting from each approach are profound. For example, in the low diffusivity scenario\footnote{Typically $\eta = 10^{6}-10^{12}\SI{}{\square\centi\metre\per\second}$ from bottom to top of the convection zone \citep{dikpati2006simulating}, produce diffusion time across the convection zone of 200 years.} where advection is the main mechanism communicating the magnetic field in the majority of the convection zone, a cycle memory effect exists in dynamo models due to high latency in the magnetic field communication. On the other hand, high diffusivity favours faster communication and interaction between magnetic field components, facilitating the establishment of hemispherical dipolar parity and restraining hemispheric asymmetry. 

At present the high turbulent diffusivity approach has shown reasonable results in reproducing solar cycles behaviours as discussed above implying that long-term solar activity memory, may be unnecessary to explain the behaviour of solar cycles. Furthermore, the observed lack of correlation between solar activity during a given Schwabe cycle and the polar magnetic field observed by its end~\citep{jiang2007solar}, highlight the randomness involved in the creation of the polar field and thereby the relevance of solar polar magnetic precursor forecasts. The high magnetic diffusivity approach used in this investigation is similar to that reported by various authors~\citep{chatterjee2004full,hazra2014deep,jiang2007solar,karak2010importance} with separated poloidal and toroidal diffusivity components in equation~\eqref{etapprof} and equation~\eqref{etatprof}\footnote{This equation differs in $0.95 R_{\sun}$ reported in the reference model and uses $0.975 R_{\sun}$ instead, errata reported in \citep{choudhuri2005user}}. Suitable diffusivity model parameter values, in combination with aforementioned solar dynamo parameterisations, are obtained by solar dynamo calibration with observed solar activity patterns, e.g. sunspot butterfly diagrams. In this investigation this has led to the use of $\eta_{SCZ1} = \SI{3e10}{\square\centi\metre\per\second}$, a slightly lower value than the reference model $\eta_{SCZ1} = \SI{4e10}{\square\centi\metre\per\second}$, and similar $\eta_{RZ} = \SI{2.2e8}{\square\centi\metre\per\second}$ and $\eta_{SCZ} = \SI{2.4e12}{\square\centi\metre\per\second}$ to the reference model. 
\begin{equation}\label{etapprof}
\eta_p = \eta_{RZ} + \frac{\eta_{SCZ}}{2}\left[1 + \mathrm{erf}\, \left( \frac{r - 0.70 R_{\sun}}{0.025 R_{\sun}}  \right) \right]
\end{equation}
\begin{equation} \label{etatprof}
\begin{split}
\eta_t = &~ \eta_{RZ} + \frac{\eta_{SCZ1}}{2}\left[1 + \mathrm{erf}\, \left( \frac{r - 0.72 R_{\sun}}{0.025 R_{\sun}}  \right) \right] + \\
& \frac{\eta_{SCZ}}{2}\left[1 + \mathrm{erf}\, \left( \frac{r - 0.975 R_{\sun}}{0.025 R_{\sun}}  \right) \right]
\end{split}
\end{equation}
It is expected that the strong toroidal field at the bottom of the convection zone, where it is supposedly generated by high differential rotation, presents suppressed diffusivity. This is a plausible assumption because dynamically stable plasma is found in the near-radiative zone and because strong flux tubes may contribute to reduce turbulent diffusivity. Additionally, the extended suppressed toroidal diffusivity in this region, helps to preserve the former high-latitude toroidal magnetic field whilst it is transported frozen to the moving plasma by the equatorward counter flow (meridional circulation near the bottom of the convection zone). As the toroidal field is stronger than the poloidal field within the convection zone, lower diffusivity in the toroidal field is expected until it matches the estimated common diffusivity at the surface. Unlike the double step profile of $\eta_t$, the weak poloidal field may be easily diffused throughout the convection zone suggesting a single step profile.

\subsubsection{Solar Activity Simulation}\label{sec:sas}

\begin{figure*}
\begin{center}
  \includegraphics[width=0.691\textwidth]{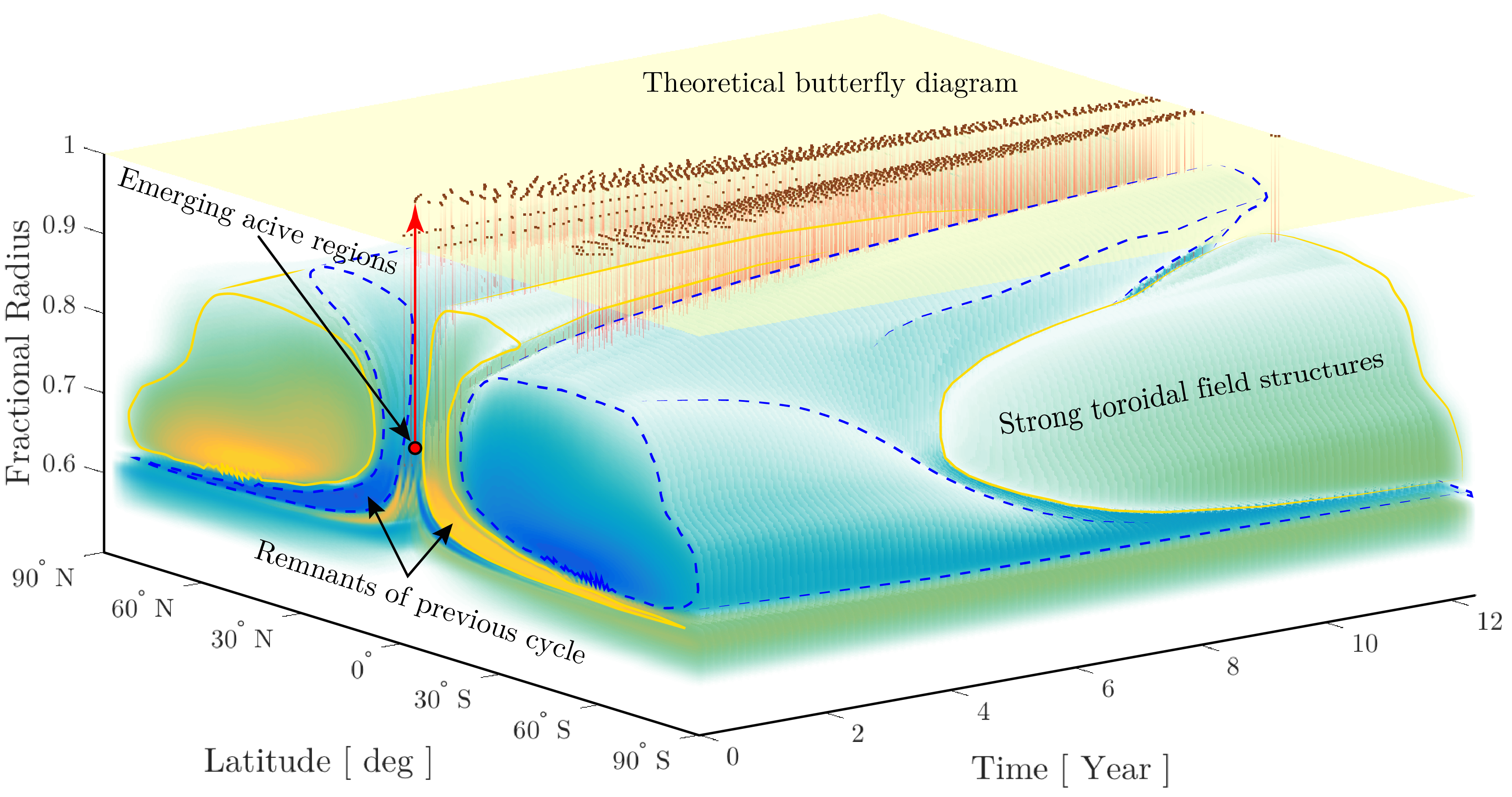}
  \end{center}
  \caption{Theoretical toroidal evolution of strong magnetic field. Weak magnetic field regions, of less than approximately $\unit[10^3]{G}$, have been removed for illustration purposes. Light solid and dark dashed lines represent magnetic field of opposite polarity. Active regions correspond to magnetic field strengths higher than $\unit[10^5]{G}$, which are believed to be the source of buoyant rise of magnetic flux tubes generating sunspots. In this illustration, emerging active regions are projected on the surface plane constituting a theoretical butterfly diagram.}
\label{fig:torosunevol}
\end{figure*}
\begin{figure}
\begin{center}
  \includegraphics[width=0.45\textwidth]{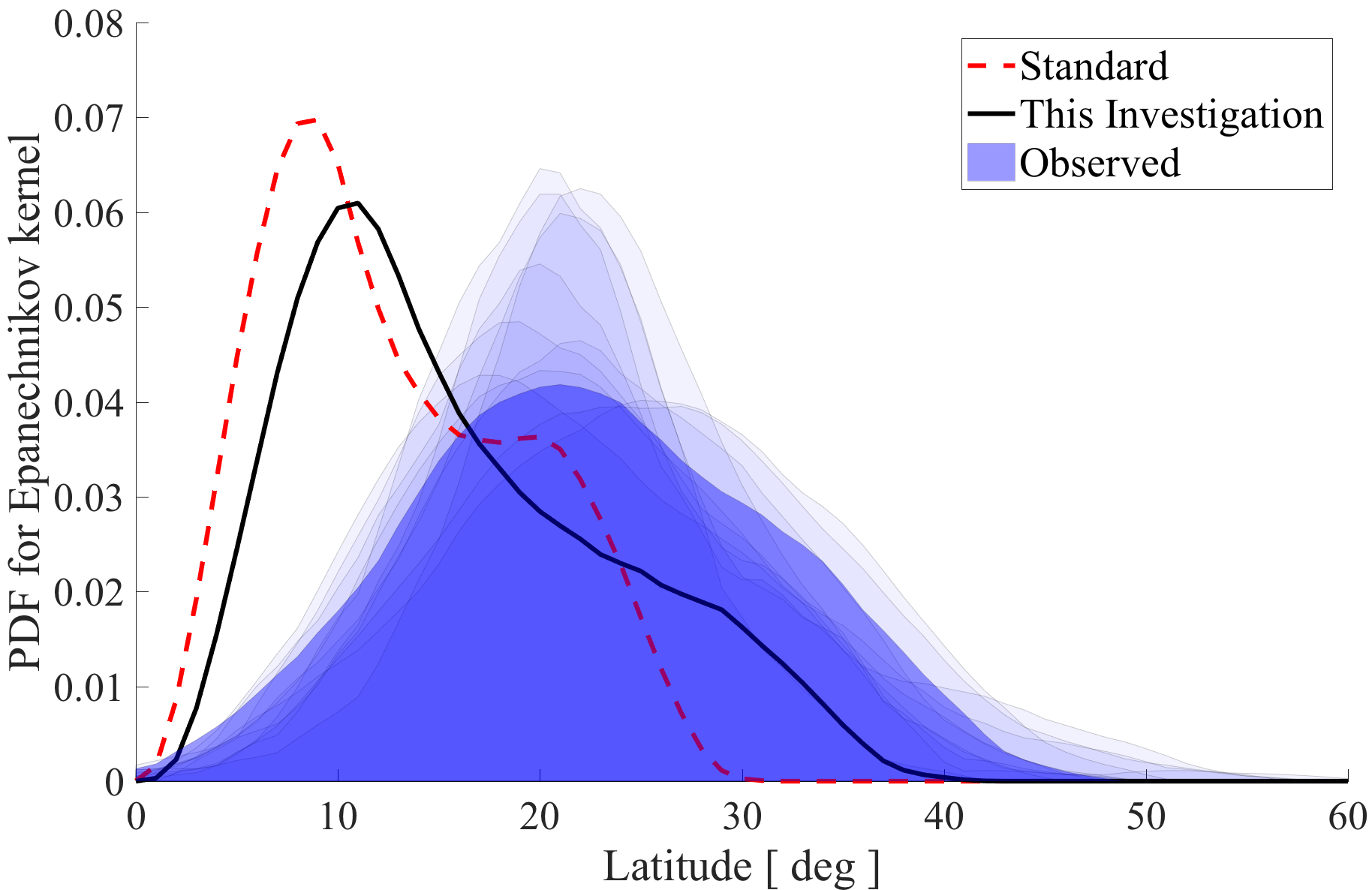}
  \end{center}
  \caption{Probability density functions from emerging active regions and regular sunspot areas. The curves represent the simulated average solar cycle activity, superimposed shaded areas correspond to observed solar cycles. The darkest shaded area corresponds to the solar cycle 23 which importance for this investigation is discussed in section \ref{sec:solarcyfore}.}
\label{fig:comparison}
\end{figure}

Sunspots, regions subject to magnetic buoyancy, are considered as a suitable proxy of solar activity for comparison between observational and simulated data. In this respect, the magnetic growth suppression approach implemented in the SURYA algorithm is used. In this approach the condition of the toroidal field at $0.71 R_{\sun}$ is monitored every 10 simulation days~\citep{choudhuri2005user}. Wherever the field exceeds the threshold of $\unit[10^5]{G}$ in this layer, half of this field is transferred to upper grid points to emulate a magnetic buoyancy-like effect. Thereby these grid points represent active regions where sunspots may appear. Figure~\ref{fig:torosunevol} shows the mean toroidal field component evolution in a steady dynamo solution (super-critical condition) from the code used in this investigation. The figure shows the characteristic toroidal field evolution during a simulated Schwabe cycle wherein weak field regions have been removed for illustration purposes. The complimentary Video~\ref{fig:suppmaterial} shows in detail the evolution of the toroidal field strengths and highlights the ultimate differences between the standard parameterisations, and those from this investigation. Note that the solar dynamo simulation using the aforementioned settings corresponds to an average solar cycle and that the number of active regions is defined by the resolution of the simulation, the implications of these factors are discussed in section \ref{sec:solarcyfore}; in this investigation a meridional slab with 256 radial and 256 angular partitions is used.

Figure~\ref{fig:comparison} presents the probability density functions (PDF) of emerging active regions during a simulated Schwabe cycle from both standard parameterisations and those used in this investigation. The superimposed shaded areas correspond to the PDFs from observed regular sunspot area activity (greater than 100 millionths of a hemisphere) during the cycles $12-23$. The set of parameterisations presented in this investigation favour the growth of azimuthal flux in a more even fashion at mid-latitudes as seen from the comparison of the right tail of the PDFs. In addition to this, the skewness of the PDF of this investigation shows better resemblance to the PDF from observational data. Assuming that magnetic tension generated by a strong toroidal field at the bottom of the convection zone is able to restrain or even momentarily stop its advection by the equatorward counterflow, the incorporation of this phenomenon in the dynamo model could favour the release of solar activity at higher latitudes further improving the PDF profile. However, further work is required to verify this hypothesis, as such in this investigation we use the dynamo parameterisations described above. The PDFs correlation coefficients between observed and simulated solar activity are 0.4921 and 0.6374 for the standard parameterisations and those used in this investigation respectively. Thus the parameterisations in the current study represent an improvement over the standard model.

\section{Solar Cycles Forecasting}
\label{sec:solarcyfore}

The designed solar-like dynamo model reaches a steady state activity response\footnote{The initial magnetic state used in this investigation is zero polar field in both hemispheres and an arbitrary toroidal magnetic field strength of equal but opposite polarity for hemispheres to establish dipolar parity. Although dipolar parity is observed in the recent record of solar activity, it is argued that quadrupolar parity may appear under specific conditions, e.g. superimposed beating magnetic configuration components. Analysis of J.C. Staudacher's sunspot drawings covering the period $1749-1796$ made by~\citet{arlt2009butterfly} and further studied by~\citet{sokoloff2009sunspot} in the context of solar dynamo theory, suggest that cycles 0 and 1 may have shown quadrupolar parity.} from an arbitrary initial magnetic state after a small number of simulated Schwabe cycles. It is worth mentioning that the stable characteristic response and the duration of the transient response during the initialisation of the dynamo model depend on the parameterisations employed. Since parameterisations remain invariant with time in the dynamo model, and these represent well-known average data in the case of the differential rotation and to a lesser extent in the meridional circulation, the simulated solar activity corresponds to an average solar cycle. In reality, fluctuations in all solar parameters confer uniqueness to each observed solar cycle. For this reason, current axisymmetrisation of the solar problem implemented in the solar dynamo, and poorly understood internal phenomena such as the possible occurrence of velocity quenching in the equatorward counterflow at the base of the convection zone, lead to coarse albeit representative approximations of observed solar activity. In addition to this, as discussed in the previous subsection \ref{sec:sas}, the number of active regions is defined by the resolution of the simulation, that is the denser the meridional slab grid the higher the number of active regions although the activity pattern evolution is preserved. For these reasons, the following effort is directed towards the identification of an observed average solar cycle to suitably scale and fine tune the dynamo model output.

\begin{figure*}
  \begin{center}
      \includegraphics[height=3.2in]{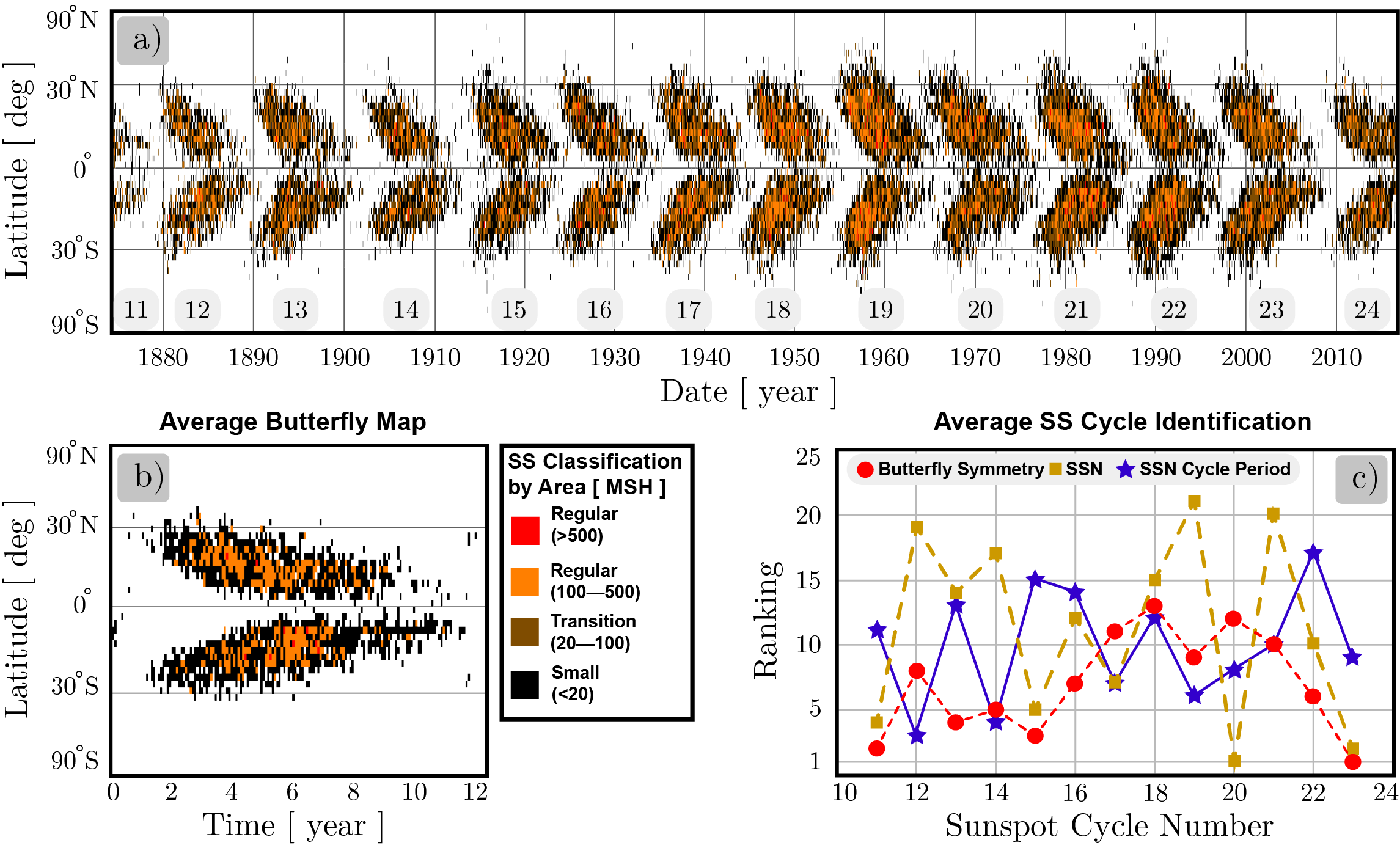}
    \caption{The average solar cycle. Subfigure $a)$ presents the record of hemispheric sunspot areas as function of solar latitude and time. Subfigure $b)$ shows the average butterfly map computed with data from the observed record. This benchmark map assist the ranking of the solar cycles, wherein rank 1 corresponds to closest similarity to this benchmark. Data from the SSN record provides two ranking benchmarks, the cycle period and SSN magnitude. The cycle number intersection of the three ranking sets is shown in Subfigure $c)$. According to the cumulative ranking from the three indicators, solar cycle 23 is identified as the average cycle.}
    \label{fig:realsynop}
  \end{center}
\end{figure*}

In order to identify the average solar cycle, in this investigation the total sunspot area record reported in the butterfly diagrams\footnote{\label{BFDfootnote}\url{https://solarscience.msfc.nasa.gov/greenwch.shtml}} and the monthly average SSN record\footnote{\label{SSNfootnote}\url{http://www.sidc.be/silso/datafiles}} are assessed. The SSN record encompasses the last 23 full cycles and the butterfly diagrams, containing hemispheric activity details shown Figure~\ref{fig:realsynop}a, the last 12 full cycles. From the SSN record, cycle periods and maximum activity magnitudes are evaluated, whilst from the hemispheric sunspot area record latitudinal activity is evaluated. In this investigation hemispheric sunspot area data (in units of millionths of a hemisphere MSH) is grouped into small sunspots, transition sunspots, and regular sunspots according to the classification presented by~\citet{tlatov2013bimodal}. By processing the complete record of hemispheric sunspot areas, it is possible to identify average sunspot activity as function of solar latitude and solar cycle evolution as shown in Figure~\ref{fig:realsynop}b. The resulting average butterfly map is then used to quantify differences in hemispheric activity for each solar cycle. The cycle with closest hemispheric activity similarity to the average butterfly map is assigned with rank 1, up to the least similar with rank 13 in this case, i.e. each cycle in the butterfly record is related to a rank value. In a similar way, the identification of the average Schwabe period and SSN magnitude from the SSN record assists the ranking process of the solar cycles. Figure~\ref{fig:realsynop}c summarises the data assessment for the common cycles and identifies the solar cycle 23 as the average cycle according to the cumulative ranking from the three indicators, i.e. butterfly diagrams, SSN maxima, and SSN cycle periods.

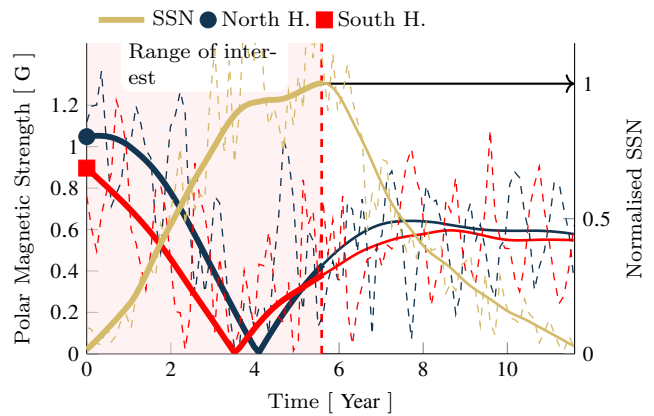
\begin{figure}
 \begin{center}
\begin{tikzpicture}
\pgfplotsset{width=8cm, height=5.7cm}
\begin{axis}[x tick label style={/pgf/number format/1000 sep=},
    xlabel={$y$},
    xmin=0,
    xmax=11.6,
    ymax=1.5, ymin=0,
    ytick={0.0,0.2,0.4,0.6,0.8,1.0,1.2},
    xlabel = {Time [  $\unit[]{Year}$ ]},
    ylabel = {Polar Magnetic Strength [  $\unit[]{G}$ ]},
    axis x line*=bottom,
    axis y line*=left,
    xmajorgrids=false,
    extra x tick style={grid=major, grid style={dashed,black}}, 
    major grid style={black!20}
]
\fill[color=red!10,opacity=0.5] (axis cs:0,0) rectangle (axis cs:5.5852,5);
\node[text width=2cm, fill=lightgray!0, rounded corners] at (axis cs:2.8,1.4) {Range of interest};
\addplot[forget plot,color= red,dashed, line width = 1pt, mark=none] coordinates{(5.5852,0) (5.5852,2)};
\addplot[forget plot,color=mylilas,line width=1pt,solid] table[x index=0,y index=3,col sep=comma,mark=none] {d_sunobserved.dat};
\addplot[forget plot,color=red,line width=1pt,solid] table[x index=0,y index=4,col sep=comma,mark=none] {d_sunobserved.dat};
\addplot[color=mylilas,line width=0.5pt,dashed] table[x index=0,y index=1,col sep=comma,mark=none] {d_sunobserved.dat};
\addplot[forget plot,color=mylilas,line width=2.2pt,solid] table[x index=0,y index=3,col sep=comma,mark=none,skip coords between index={49}{100}] {d_sunobserved.dat};
\addplot[forget plot,color=red,line width=0.5pt,dashed] table[x index=0,y index=2,col sep=comma,mark=none] {d_sunobserved.dat};
\addplot[forget plot,color=red,line width=2.2pt,solid] table[x index=0,y index=4,col sep=comma,mark=none,skip coords between index={49}{100}] {d_sunobserved.dat};
\addplot [color=mylilas, only marks,mark size=3pt,mark=*] coordinates{(0,1.0484)};\label{plot1obs}
\addplot [color=red, only marks,mark size=3pt,mark=square*] coordinates{(0,0.89678)};\label{plot2obs}
\end{axis}
\begin{axis}[x tick label style={/pgf/number format/1000 sep=},
    color=black, 
    xmin=0,
    xmax=11.6,
    ymin = 0,
    ymax = 1.15,
    axis y line*=right,
    ylabel = {Normalised SSN},
    ylabel near ticks, yticklabel pos=right,
    hide x axis,
    legend style = {at={(0.725,1.01)}, anchor = south east, legend columns = -1, draw=none, font=\small}
]	
\draw[thick,->] (axis cs:5.5852,1) -- (axis cs:11.6,1);
\addplot[forget plot,color=mygreen,line width=1pt,solid] table[x index=0,y index=6,col sep=comma,mark=none] {d_sunobserved.dat};
\addplot[color=mygreen,line width=2.2pt,solid] table[x index=0,y index=6,col sep=comma,mark=none,skip coords between index={49}{100}] {d_sunobserved.dat};\label{plot3obs}
\addplot[forget plot,color=mygreen,line width=0.5pt,dashed] table[x index=0,y index=7,col sep=comma,mark=none] {d_sunobserved.dat};
\addlegendimage{refstyle=plot1obs} \addlegendentry{SSN}
\addlegendimage{refstyle=plot2obs} \addlegendentry{North H.}	
\addlegendimage{refstyle=plot3obs} \addlegendentry{South H.}	
\end{axis}
\end{tikzpicture}
\caption{Selected indicators for observed solar cycle 23. The dark solid curves represent smoothed polar magnetic strength magnitudes, left scale, and the solid curve in light colour is the normalised SSN, right scale. In this investigation, the observed polar magnetic field strength values at time zero are used as modification inputs to hemispheric poloidal potential modulation in the dynamo model for each studied solar cycle. On the other hand, the observed maximum SSN value in the solar cycle 23 is used to scale the resulting theoretical dynamo activity output. In this way, observed cycle 23 SSN and pseudo-SSN maxima from the calibrated dynamo model are equivalent.}
\label{fig:suncalib2}
\end{center}
\end{figure}

Having identified the average solar cycle it is possible to establish conversion factors to relate observed solar characteristics to the dynamo model output. This fundamental procedure enables the identification of the scaling relationship between the model output, i.e. representing an average solar cycle, and the observed average solar cycle 23. Firstly, in this investigation we are interested in a conversion factor to relate the theoretical polar magnetic field strength at SSN minimum, viz. instantaneous initial state, to the corresponding observed value\footnote{\label{WSOfootnote}\url{http://wso.stanford.edu/}}; this is because the solar polar magnetic field strength is used as the main parameter for solar activity forecast in a similar approach to \citet{jiang2007solar}. The smoothed polar field, solid lines in Figure~\ref{fig:suncalib2}, represent the average observed field in the polar caps with effective latitudes at $\pm70^{\circ}$~\citep{svalgaard1978strength}. This observable surface indicator of the solar state is of central importance in the fine adjustment of the dynamo model presented in this investigation, since it is hypothesised that the magnitude of the polar magnetic field is related to the poloidal magnetic structures underneath. In this respect, the method of meridional slab modification within the region enclosed by $0.8 R_{\sun}$ and the external radius as a means to modulate the poloidal potential is implemented~\citep{jiang2007solar}. This method allows the even modulation of the poloidal magnetic field in the dynamo model slab. However, note in Figure~\ref{fig:suncalib2} that the smoothed northern and southern hemispheric polar field curves differ slightly at time zero (at SSN minimum). In order to include these observed hemispheric asymmetries, in this investigation we use a sectioned modification meridional slab divided by the solar equator. In this divided meridional slab, the magnitude of the smoothed south hemisphere polar field is chosen arbitrarily as the reference value to compute the model's scale factors; in other words the characteristic southern dynamo polar magnetic field strength at a point near the surface and solar latitude $70^{\circ}$ is proportional to the respective observed value wherein the constant of proportionality is the scale factor. Finally, scaling the theoretical emerging active regions profile resulting from the poloidal potential modulation, hereinafter referred to as pseudo-SSN, to observed solar activity is necessary; the maximum smoothed SSN is used for this purpose. Attention of subsequent analyses should be centred only on approximately the first lustrum of solar activity, e.g. the shaded area in Figure~\ref{fig:suncalib2}, corresponding to the relatively stable SSN rising cycle phase as discussed in subsection \ref{ssec:ftdt}.

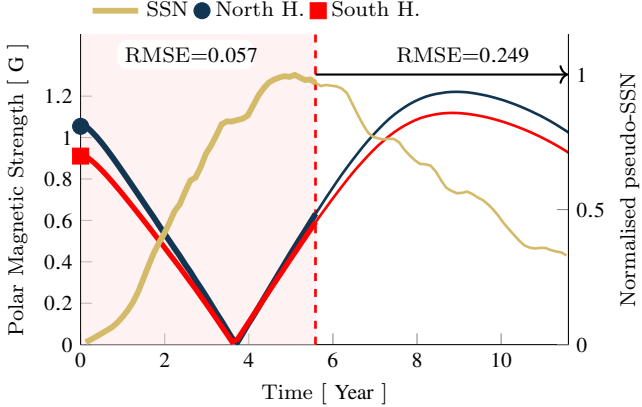
\begin{figure}
 \begin{center}
\begin{tikzpicture}
\pgfplotsset{width=8cm, height=5.7cm}
\begin{axis}[x tick label style={/pgf/number format/1000 sep=},
    xlabel={$y$},
    xmin=0,
    xmax=11.6,
    ymax=1.5, ymin=0,
    ytick={0.0,0.2,0.4,0.6,0.8,1.0,1.2},
    xlabel = {Time [  $\unit[]{Year}$ ]},
    ylabel = {Polar Magnetic Strength [  $\unit[]{G}$ ]},
    axis x line*=bottom,
    axis y line*=left,
    xmajorgrids=false,
    extra x tick style={grid=major, grid style={dashed,black}}, 
    major grid style={black!20}
]
\fill[color=red!10,opacity=0.5] (axis cs:0,0) rectangle (axis cs:5.5852,5);
\addplot[forget plot,color= red,dashed, line width = 1pt, mark=none] coordinates{(5.5852,0) (5.5852,2)};
\node[text width=1.6cm, fill=lightgray!0, rounded corners] at (axis cs:2.5,1.4) {RMSE=$0.057$};
\node[text width=2.2cm, fill=lightgray!0, rounded corners] at (axis cs:9.5,1.4) {RMSE=$0.249$};
\addplot[forget plot,color=mylilas,line width=1pt,solid] table[x index=0,y index=1,col sep=comma,mark=none] {d_sundynamo.dat};
\addplot[forget plot,color=red,line width=1pt,solid] table[x index=0,y index=2,col sep=comma,mark=none] {d_sundynamo.dat};
\addplot[forget plot,color=mylilas,line width=2.2pt,solid] table[x index=0,y index=1,col sep=comma,mark=none,skip coords between index={45}{100}] {d_sundynamo.dat};
\addplot[forget plot,color=red,line width=2pt,solid] table[x index=0,y index=2,col sep=comma,mark=none,skip coords between index={45}{100}] {d_sundynamo.dat};
\addplot [color=mylilas, only marks,mark size=3pt,mark=*] coordinates{(0,1.0543)};\label{plot1sim}
\addplot [color=red, only marks,mark size=3pt,mark=square*] coordinates{(0,0.91055)};\label{plot2sim}
\end{axis}
\begin{axis}[x tick label style={/pgf/number format/1000 sep=},
    color=black, 
    xmin=0,
    xmax=11.6,
    ymin = 0,
    ymax = 1.15,
    axis y line*=right,
    ylabel = {Normalised pseudo-SSN},
    ylabel near ticks, yticklabel pos=right,
    hide x axis,
    legend style = {at={(0.725,1.01)}, anchor = south east, legend columns = -1, draw=none, font=\small}
]
\draw[thick,->] (axis cs:5.5852,1) -- (axis cs:11.6,1);
\addplot[color=mygreen,line width=2.2pt,solid] table[x index=0,y index=4,col sep=comma,mark=none,skip coords between index={45}{100}] {d_sundynamo.dat};\label{plot3sim}
\addlegendimage{refstyle=plot1sim} \addlegendentry{SSN}	
\addlegendimage{refstyle=plot2sim} \addlegendentry{North H.}	
\addlegendimage{refstyle=plot3sim} \addlegendentry{South H.}	
\addplot[forget plot,color=mygreen,line width=1pt,solid] table[x index=0,y index=4,col sep=comma,mark=none] {d_sundynamo.dat};
\end{axis}
\end{tikzpicture}
\caption{Same as Figure~\ref{fig:suncalib2} for the calibrated dynamo to the observed solar cycle 23. Polar field magnitudes correspond to the solar latitudes $\pm70^{\circ}$. Starting from the forced initial poloidal potential modulation to match the observed initial polar field magnitudes at time zero, the dynamo model evolves to polar field inversion around the fourth year in close resemblance to the observed trend in Figure~\ref{fig:suncalib2} within the relatively stable SSN rising phase. The maximum value of the pseudo-SSN profile, obtained from the historical number of emerging active regions, is equivalent to the maximum value of observed SSN during the identified average solar cycle 23. Bear in mind that SSN equivalence is ensured only for the solar cycle 23 whereas in other cycles it is not necessarily true as expected from the diverse polar magnetic field strengths initial inputs used in the solar dynamo.}
\label{fig:suncalib1}
\end{center}
\end{figure}

The calibrated solar dynamo model for cycle 23, that is the fully scaled dynamo output from the selected solar activity indicators, is shown Figure~\ref{fig:suncalib1}. The root-mean-square error (RMSE) of the model's pseudo-SSN and the observed SSN for the rising and declining phases are reported in the figure. The effect of forced initial hemispheric poloidal potential asymmetry in the dynamo model is clearly observed in the polar magnetic strength magnitude evolution. From the plots in Figure~\ref{fig:suncalib2} and Figure~\ref{fig:suncalib1} \textit{sufficient} resemblance during the rising phase of the SSN cycle is observed. Conversely, the most remarkable incompatibility is observed in the declining phase of the sunspot cycle as expected due to the highly random processes dominating the poloidal field generation as previously discussed. The main factor limiting conclusions about the adequacy of the solar dynamo with polar magnetic field as precursor for forecasting purposes, is the reduced record of polar solar magnetic activity encompassing only the last full three solar cycles. In order to overcome this limitation, we use extended polar field strengths reconstructed with the longer record of hemispheric sunspot areas. The reconstruction method and criteria is discussed in the following subsection \ref{sec:sopofire}. 

\subsection{Solar Polar Field Strength Reconstruction} \label{sec:sopofire}
\begin{figure}
  \begin{center}
      \includegraphics[height=1.9in]{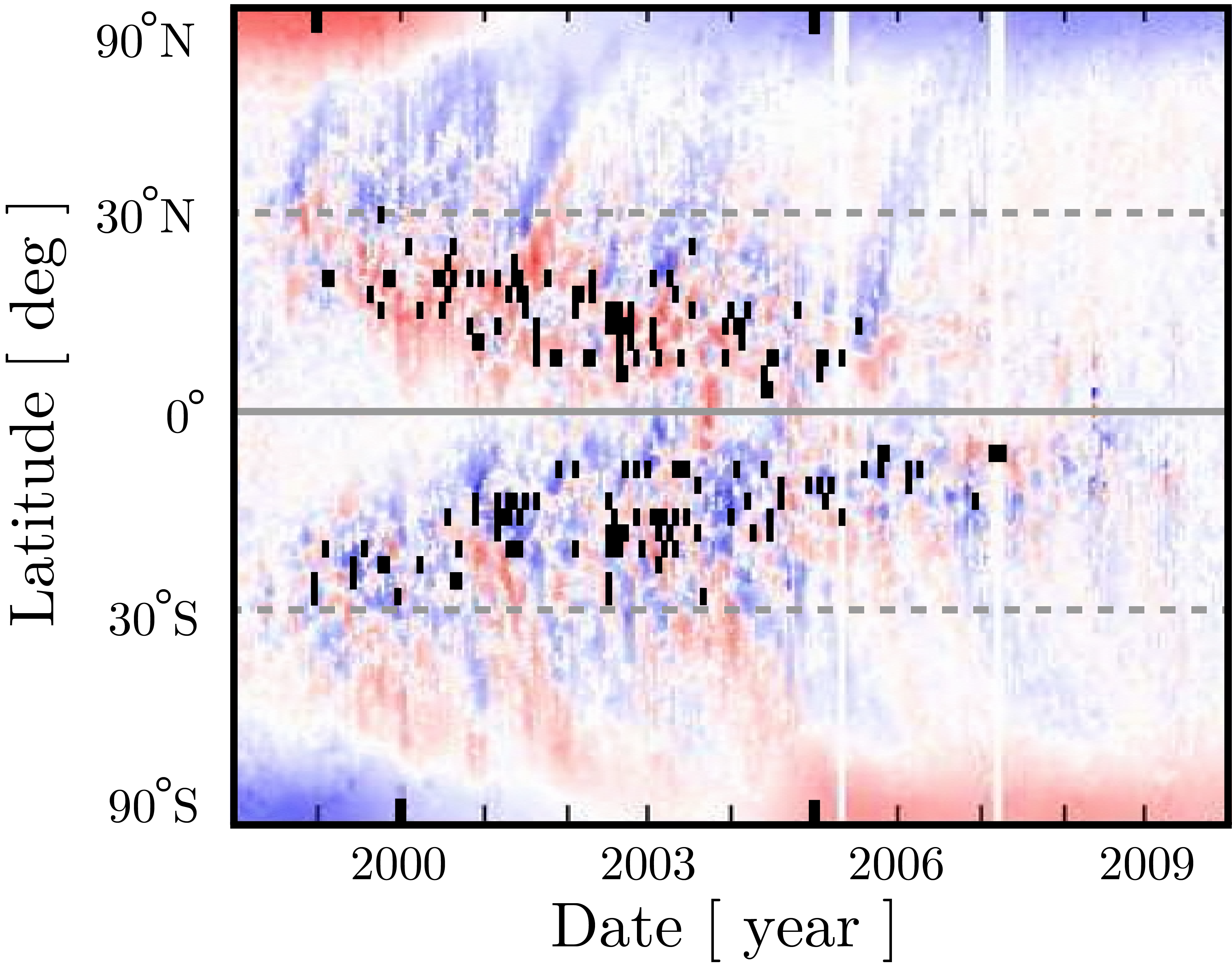}
    \caption{Longitudinally averaged magnetic field (magnetic butterfly) diagram for the solar cycle 23. Reddish colour represents positive flux, and highly active sunspot areas are marked in black. Trails of unipolar poleward drift are observed in both solar hemispheres. In this investigation is argued that regular sunspot activity is likely to be captured in the polar magnetic field pole caps as this is transported by meridional circulation. If the information reaching the pole caps contain sufficient detail about past sunspot activity, then this may open the possibility to reconstruct unrecorded historical polar field from hemispheric sunspot areas. Image adapted from \citet{petrie2013solar}}
    \label{fig:realmagsynop}
  \end{center}
\end{figure}

With the advance of the solar cycle and as the generation of poloidal flux from toroidal flux develops near the solar surface, part of the generated flux is advected by meridional flow circulation to the poles where these gradually accumulate building up a distinguishable polar field of maximum strength near the end of the Schwabe cycle. Additionally, remnants of sunspot activity, i.e. remaining flux after cancellation and submergence, is diffused locally generating unipolar patches. According to the sunspot magnetic flux level and helicity \citep{0004-637X-840-2-100}, a small amount may withstand flux fragmentation and cancelation to form long-lasting diffuse unipolar patches. These patches gradually form distinguishable trails of unipolar poleward drift with respect to the surrounding average weak diffuse magnetic field; these characteristic features can be observed in the averaged photospheric magnetic field reported in synoptic magnetic butterfly diagrams, Figure~\ref{fig:realmagsynop}. However, the surface of the Sun is intrinsically active where magnetic field lines are in continuous efficient interaction. Spontaneous magnetic field lines emerging throughout the surface are fragmented, redistributed, and diffused in the magnetic carpet renewing itself in a few tens of hours ($10-40$ hours~\citep{priest2014magnetohydrodynamics}). This inherent surface activity affects trails of unipolar regions enhancing their decline. Thus, regular sunspots (flux density above $\sim\unit[10^{21}]{Mx}$~\citep{tlatov2013bimodal}) are more likely to contribute to long-lasting unipolar patches than small sunspots or pores ($\lessapprox \unit[10^{20}]{Mx}$). For this reason, in this investigation it is assumed that the information about past solar activity conveyed to the pole caps is plausibly related to chiefly regular sunspot activity. The overall aforementioned process generates a delay in the evolution of the polar magnetic field with respect to the sunspot activity. In this investigation, the phase shift between both proxies is identified at 1500 days (4.1 years) via cross-correlation analysis of the SSN\textsuperscript{\ref{SSNfootnote}} record and the Wilcox Solar Observatory polar magnetic field record (WSO)\textsuperscript{\ref{WSOfootnote}}. On the grounds of this reasoning, regular sunspots in the sunspot areas record (Figure~\ref{fig:realsynop}) may contain sufficient information to reconstruct unrecorded polar field cycles. Artificial neural network methods can be suitably used to address this task owing to the complex nature of the relationships involved as discussed below.

An artificial neural network (NN) is a computational model built with a collection of interconnected simple processing instances called neurones, resembling in essence a biological neural network. Each neurone consists of adjustable scalar weights to scale inputs, a fixed transfer function to model the neurone shape response, and an adjustable scalar bias to shift the function. Interconnection ensures that every neurone contributes to the collective neural network response. The objective of the NN implementation is to learn from and respond to given data in a desired way, wherein adaptability and capability of generalisation of the NN to new data is the ultimate objective. To this aim, a fundamental design stage in NN implementation is the \textit{training} phase in which the NN learns in either an unsupervised or supervised manner \citep{graupe2007principles,demuth2014neural}. During the \textit{training} phase in unsupervised NNs, the adjustable scalars in each neurone are adjusted aiming at the discrimination of data lying within specific criteria. Unsupervised NNs use input data to auto adjust their parameters dynamically. This type of NN is valuable in classification and pattern recognition, e.g. cluster analysis and image processing for object detection. On the other hand, a supervised NN uses data with labeled responses from input and expected response (target) data\footnote{One of the main differences between supervised and unsupervised NNs is that the later adjusts itself using the network inputs only since no target outputs are available.}. In this case, an iterative process modifies the adjustable scalars in each neurone during the \textit{training} phase aiming at the faithful reproduction of target reference data. Supervised NNs find application in automated classification such as face/voice recognition, and in the modelling of continuous system response as a function of input predictor variables (regression). This later capability of supervised NNs is typically applied in time series predictions such as in financial analyses or even proposed for SSN trend prediction \citep{li1990forecasting}. For the problem at hand, supervised NN training under the regression category was selected because the training dataset, i.e. predictor and target data, encompass sequential vectors.

When dealing with sequential vectors as NN inputs, adding feedback and delays typically improves the NN performance because these establish memory capability valuable for learning sequential patterns. This type of NN model is called a dynamic network, and relates the history of the input sequence and feedback sequence to establish the NN response. A Nonlinear AutoRegressive network with eXogenous inputs (NARX) is a dynamic model that can use several layers of neurones and independent input series, namely \textit{exogenous series}, which are known to influence the target series. The NARX model output $y \left( t \right)$ is thus a function of past values of $y\left( t - n_y \right)$ (target or feedback data), and exogenous series $x\left( t - n_x\right)$, Eq.~\eqref{E_narxnet}. The powerful NARX model may be applied in a myriad of applications. In this investigation the NARX model is used to represent a nonlinear dynamic system relating present and past data of the driving hemispheric sunspot area series (exogenous), and the record of solar polar field (target). The selection of this input dataset for the NARX model is justified due to the known influence of the exogenous series on the target series as discussed earlier in this section.
\begin{equation}\label{E_narxnet}
y \left( t \right) = f \left[ y\left( t - 1\right), \dotsc, y\left( t - n_y \right), x\left( t - 1\right), \dotsc, x\left( t - n_x\right)\right]
\end{equation}

 \begin{figure*}
  \begin{center}
      \includegraphics[height=1.7in]{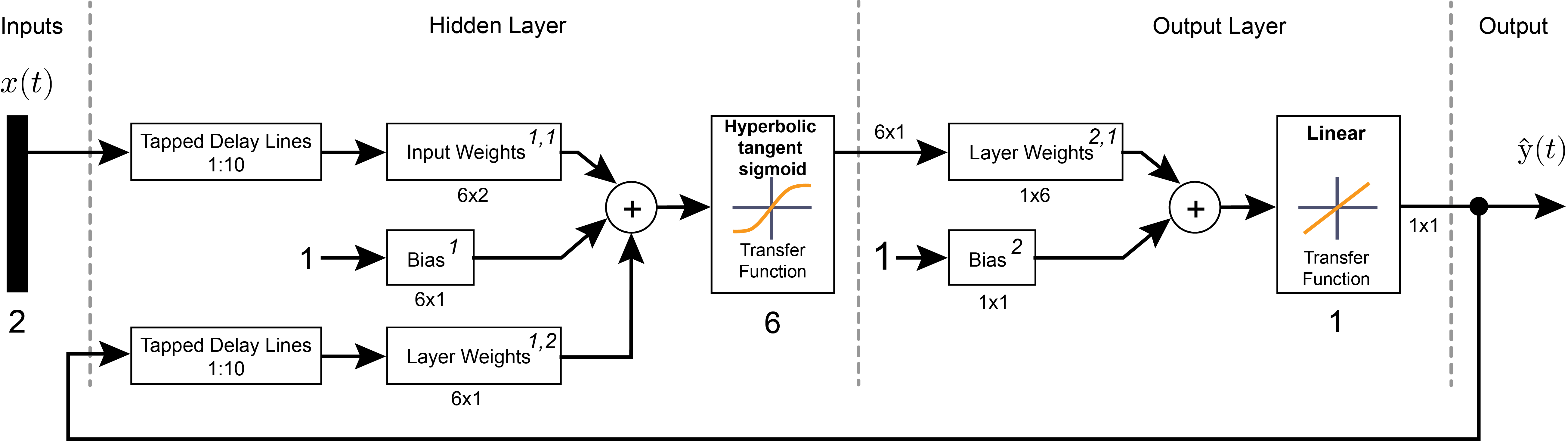}
    \caption{Layer diagram of the Nonlinear AutoRegressive network with eXogenous inputs (NARX) in parallel configuration (closed loop). The network uses 6 hidden layer neurones and the input and feedback delays range $1-10$. The two exogenous inputs correspond to the hemispheric regular sunspot area records. During the training phase, that is the procedure to establish weights and biases values, the target series (one observed hemispheric polar field series at a time) is available requiring a series-parallel architecture; after the training phase, model output feedback is used in replacement of the target series in the parallel architecture shown. Architecture change calls for further generalisation performance assessment to discard possible target memorisation during the training phase.}
    \label{fig:narxnet}
  \end{center}
\end{figure*}

The NARX model used for the reconstruction of the polar field in this investigation is presented in Figure \ref{fig:narxnet}. The two exogenous input series ($x_{1,2}\left(t \right)$) are the north and south hemispheric regular sunspot areas, and the target series ($y\left(t \right)$) is an observed polar field. It should be noted that the two hemispheric exogenous input series are used in the proposed reconstruction of each solar polar field. This is a reasonable approach assuming that sufficient magnetic communication exists between both hemispheres as suggested by the nearly mirrored polar magnetic strength history in Figure~\ref{fig:suncalib2}. Hence, possible loss of past solar activity information during the creation of unipolar patches and the subsequent hemispheric poleward drift, may be plausibly recovered from the regular sunspot areas record of both hemispheres. Resuming the description of the model, a layer of six interconnected sigmoid neurones in the so-called hidden layer, are fed with the exogenous and target series delayed $1-10$ states ($10$ Julian days per state). This delay range is heuristically established according to the ultimate performance of the model. Finally, the output layer consists of a linear neurone used as a function approximator.

\begin{figure*}
 \begin{center}
\begin{tikzpicture}[spy using outlines={circle, magnification=6, size=2.3cm, connect spies}, thick,scale=1.0]
\pgfplotsset{width=17cm, height=6.9cm}
\begin{axis}[x tick label style={/pgf/number format/1000 sep=},
    no markers,grid=major, 
    xmin=1975, xmax=2027,
    ymin=-3500, ymax=3500,
    xtick={1980,1990,2000,2010,2020},
    axis lines = left,
    xlabel = {Date [ Year ] },
    ytick=\empty,
    ylabel = {Series (Exogenous / Polar Field) [ MSH / G ] },
    extra y tick style={grid=major, grid style={dashed,black}}, 
    major grid style={black!20},
    legend style = {at={(0.925,1.01)}, anchor = south east, legend columns = -1, draw=none, font=\small}
]
\fill[color=red!10,opacity=0.5] (axis cs:2015,-50000) rectangle (axis cs:2017,50000);
\addplot[forget plot,color= gray,dashed, line width = 1pt, mark=none] coordinates{(2015,-50000) (2015,50000)};	
\addplot[forget plot,color= gray,dashed, line width = 1pt, mark=none] coordinates{(2017,-50000) (2017,50000)};
	
\addplot[color=orange,line width=1pt,line join=round] table[x index=0,y index=1,col sep=comma,mark=none] {d_n_referencevect.dat};
\draw[thick,->] (axis cs:1980,2000) -- (axis cs:1984.5,2000);
\draw[thick,->] (axis cs:1980,-2000) -- (axis cs:1984,-2000);
\node[text width=0.6cm, fill=lightgray!20, rounded corners] at (axis cs:1980,2000) {$x_1\left(t\right)$};
\node[text width=0.6cm, fill=lightgray!20, rounded corners] at (axis cs:1980,-2000) {$x_2\left(t\right)$};

\addplot[color=mylilas,line width=1pt,line join=round,dotted] table[x index=0,y index=2,col sep=comma,mark=none] {d_n_referencevect.dat};
\addplot[color=mygreen,line width=2.5pt,line join=round] table[x index=0,y index=3,col sep=comma,mark=none] {d_n_referencevect.dat};
\addplot[forget plot,color=mylilas,line width=1pt,line join=round,dotted] table[x index=0,y index=2,col sep=comma,mark=none] {d_n_forecastvect.dat}; \addlegendentry{Exogenous input series}
\addplot[forget plot,color=orange,line width=1pt,line join=round,solid] table[x index=0,y index=1,col sep=comma,mark=none] {d_n_forecastvect.dat}; \addlegendentry{Exogenous input series}
\addplot[forget plot,color=mygreen,line width=2.5pt,line join=round] table[x index=0,y index=1,col sep=comma,mark=none] {d_n_forecas2vect.dat};\addlegendentry{Target series}
\addplot[color=blue,line width=2pt,line join=round] table[x index=0,y index=3,col sep=comma,mark=none] {d_n_forecastvect.dat};\addlegendentry{Estimated series}
\draw[{Circle[red]}-Latex] (axis cs:2016,3000) -- (axis cs:2020,3000);
\node[text width=1.7cm, fill=lightgray!20, rounded corners] at (axis cs:2005,3000) {Training phase};
\node[text width=1.3cm, fill=lightgray!20, rounded corners] at (axis cs:2022,3000) {Evaluation phase};
\draw[thick,->] (axis cs:2000,-2000) -- (axis cs:1995,-190);
\node[text width=0.55cm, fill=lightgray!20, rounded corners] at (axis cs:2000,-2000) {$y\left(t\right)$};
\draw[thick,->] (axis cs:2021,-2000) -- (axis cs:2016.06,-190);
\node[text width=0.55cm, fill=lightgray!20, rounded corners] at (axis cs:2021,-2000) {$\hat{y}\left(t\right)$};
\coordinate (spypoint) at (axis cs:2016,-140);
\coordinate (magnifyglass) at (axis cs:2010,-2000);
\end{axis}
\spy [blue, size=2.2cm] on (spypoint)
   in node[fill=white] at (magnifyglass);
\end{tikzpicture}
\caption{Time series employed for the southern polar field short-term forecast. In this example all relevant time series are superimposed for illustrative purposes only. Phase shift between observed hemispheric sunspot areas (exogenous inputs) and polar field strength (target) is cancelled during computations. Multiple NARX models are trained to finally select one able to generalise best during the evaluation phase. The magnified area shows the target series segment $y\left(t\right)$ and the estimated value of the polar field $\hat{y}\left(t\right)$. The best NARX model is expected to represent a nonlinear dynamic system relating exogenous input series and target series. In total, a polar magnetic strength forecast and a reconstruction is obtained for each hemisphere.}
\label{fig:neuralsample}
\end{center}
\end{figure*}
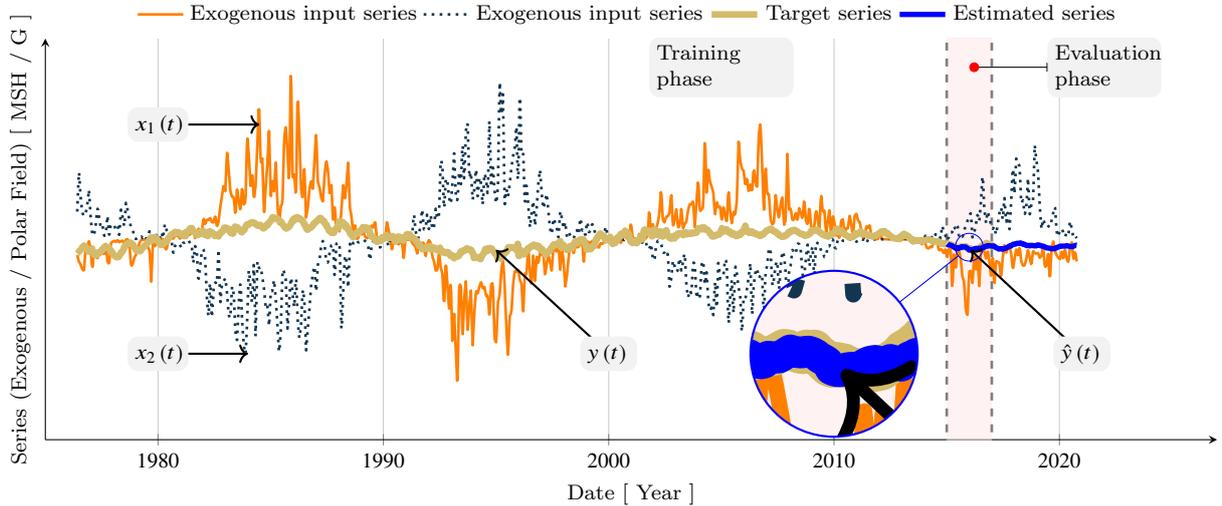

Specifically, Figure~\ref{fig:narxnet} illustrates a parallel NARX architecture (closed loop) useful in the practical application of the model. In a previous step, the model is trained to establish suitable weights and bias values. During the training phase the target series, in this case a polar field series, is available requiring no $\hat{y}\left(t \right)$ feedback. The training algorithm adopted is the standard Levenberg-Marquardt backpropagation, with internal data division of $70\%$ training, $15\%$ validation, and $15\%$ testing. During the training phase, the Levenberg-Marquardt optimisation looks for adequate weights and bias values to reduce the error between the target signal and the model's output. Since internal data division and sampled data for training is stochastic, mainly to prevent target memorisation, computed weights and biases may differ amongst various trained models leading to various performance levels. This means that a trained model would have a unique set of bias and weight values related to the \textit{randomly} sampled input data required by the training algorithm. Furthermore, depending on the overall sampled data during the training phase, the NARX may be more or less adaptable to interpret \textit{new} input data according to its success in building a representative nonlinear dynamic system. Aiming at improving NN generalisation, multiple models may be trained to select the one able to generalise best. In this investigation 200 models for each polar field are trained, and in order to identify the most accurate one, the available target data is divided in two parts: $95\%$ is used for the model training phase, and the remaining $5\%$ is saved for a subsequent best model identification phase. The optimum trained NARX model for each hemisphere is used to reconstruct the respective polar field. In this implementation phase, the target input $y\left(t \right)$ is substituted by $\hat{y}\left(t \right)$ transforming the series-parallel architecture (open loop) into the parallel architecture shown in Figure~\ref{fig:narxnet}. The regular sunspot area record provides sufficient exogenous input series to reconstruct polar field from the Schwabe cycle 12 to 20, and to forecast their values up to the next sunspot minimum. As mentioned before, this last feature is possible due to the the existing phase shift between the sunspot area record and polar field. In Figure~\ref{fig:neuralsample} the exogenous input series, target series, estimated series, and data segmentation for the training and best model identification phases are identified. 

\begin{figure*}
 \begin{center}
\begin{tikzpicture}
\pgfplotsset{width=17cm, height=7.2cm}
\begin{axis}[x tick label style={/pgf/number format/1000 sep=},
    grid=major, 
    xmin=1877, xmax=2027,
    ymin=0, ymax=2.2,
    axis lines = left,
    xlabel = {Date [ Year ] },
    ylabel = {Polar Field Strength [ $\unit[]{G}$ ]},
    ymajorgrids=true,
    extra y tick style={grid=major, grid style={dashed,black}}, 
    major grid style={black!20},
    legend style={legend pos=north west, font=\small, legend cell align=left}
]
\fill[color=red!10,opacity=0.6] (axis cs:1976.418562660025,0) rectangle (axis cs:2017.018836445103,2);
\node[right,text width=1.3cm, fill=lightgray!0, rounded corners] at (axis cs:1990,1.8) {WSO Data}; 
\node[right,text width=0.8cm, fill=lightgray!20, rounded corners] at (axis cs:1955,1.8) {NARX}; 
\node[right,text width=0.8cm, fill=lightgray!20, rounded corners] at (axis cs:2017,1.8) {NARX}; 

\addplot[color=mylilas,line width=1pt] table[x index=0,y index=1,col sep=comma,mark=none] {d_wso_recav.dat};
\addlegendentry{North Hemisphere}
\addplot[color=mygreen,line width=1pt] table[x index=0,y index=2,col sep=comma,mark=none] {d_wso_recav.dat};
\addlegendentry{South Hemisphere}
\addplot[only marks, mark=square*,mark size=3pt,fill=gray!20] plot coordinates{
(2050,0)};
\addlegendentry{Ref. States for S. Dynamo}

\addplot[only marks,color=blue,mark=triangle*,mark size=5pt] table[x index=0,y index=1,col sep=comma] {d_faculae.dat};
\addlegendentry{Avg. PF Maxima Date \cite{munoz2012calibrating}}

\addplot[color=mygreen,mark=square*,only marks,mark size=3pt] table[x index=0,y index=2,col sep=comma] {d_wso_sim.dat};
\addplot[color=mylilas,line width=0.8pt,loosely dotted] table[x index=0,y index=1,col sep=comma,mark=none] {d_wso_rec.dat};
\addplot[color=mygreen,line width=0.8pt,loosely dotted] table[x index=0,y index=2,col sep=comma,mark=none] {d_wso_rec.dat};
\addplot[color=mylilas,mark=square*,only marks,mark size=3pt] table[x index=0,y index=1,col sep=comma] {d_wso_sim.dat};
\addplot[color=red,line width=1pt] coordinates {
    (1976.418562660025,0)
    (1976.418562660025,2)
};
\addplot[color=red,line width=1pt] coordinates {
    (2017.018836445103,0)
    (2017.018836445103,2)
};
\end{axis}
\end{tikzpicture}
\caption{Full solar polar field series from the NARX reconstruction and forecasts, and from the Wilcox Solar Observatory record. Triangles show dates of reported north-south maxima average polar field strength magnitude according to \citet{munoz2012calibrating}, derived from a consolidated polar faculae database. Dates of peak values from the NARX reconstruction are similar to those from the polar faculae database supporting the reconstruction validity in this respect.}
\label{fig:wsoall}
\end{center}
\end{figure*}
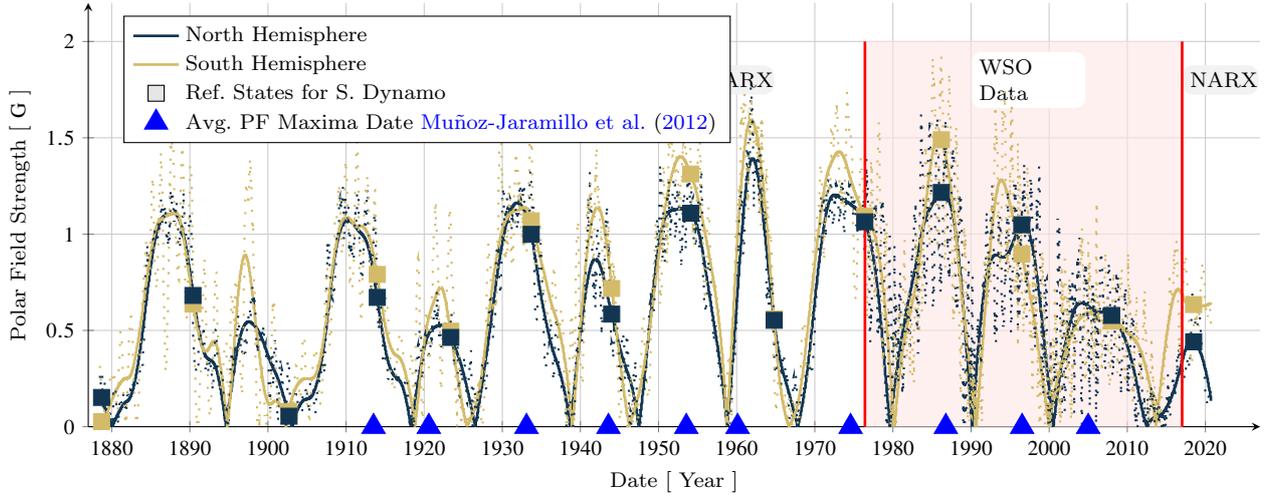

The reconstructed and forecasted polar field time series are used to establish the poloidal potential modulation in the dynamo model to reproduce observed solar activity. It is convenient at this point to bear in mind that the poloidal potential modulation is effective only at the start of each simulated solar cycle, which implies natural poloidal potential evolution along the simulated solar cycle. In this way the polar field time series are sampled at SSN minima to identify the initial poloidal potential modulation for each cycle. With the exception of the starting date of the forthcoming solar cycle 25, all other dates are established. Thus a short-term prediction to estimate the start of the solar cycle 25 is required. To this end, a standard autoregressive model of degree 9 \citep{werner2012sunspot} is used on the monthly smoothed SSN as shown in Figure \ref{fig:ssndetail}. The model short-term prediction sets the start of the solar cycle 25 around the second half of 2018\footnote{In close agreement to the SILSO\textsuperscript{\ref{SSNfootnote}} forecasts (Kalman-filter optimised forecasts ML and SC).}. In Figure \ref{fig:wsoall} the magnitude of the full solar polar field series from the NARX reconstruction and forecasts, as well as from the Wilcox Solar Observatory record are presented. Squares in the figure indicate the solar polar field values considered for subsequent solar dynamo input (at the beginning of each solar cycle). From Figure~\ref{fig:wsoall} it is observed that the NARX model methodology succeeded in obtaining a nonlinear dynamic system embodying long-term characteristics of the complex processes relating sunspot areas and polar field strengths. However cumulative short-term variations reducing polar field strength representativeness of past solar activity, such as active regions helicity hemispheric compatibility and irregular unipolar poleward migration velocity \citep{0004-637X-840-2-100}, are arguably an important source of degraded NARX model performance. An immediate apparent discrepancy observed in the NARX curves is the expected maximum of polar field magnitude at the start of the Schwabe cycles, which is true during the observed magnetic field window in the figure (WSO data). It is essential to acknowledge that the NARX curves are by no means accurate representations of polar field strength history but coarse estimations derived from a small portion of the observed Sun. However, the NARX reconstruction provides interesting clues of possible past behaviour challenging the generalisation of the aforementioned expected trends. In this regard, the consolidated polar faculae (PF) database reported by \citet{munoz2013solar} provides supporting evidence to the validity of the NARX reconstruction. For instance, the times of polar faculae maxima (marked as triangles in Figure~\ref{fig:wsoall}) although similar to the times of WSO polar field maxima and in agreement to those of SSN minima, show a local maximum near 1960 differing to the reported start of the solar cycle 20 that took place in 1964; this time of polar faculae maximum and the respective maximum from the NARX reconstruction are similar as shown in Figure~\ref{fig:wsoall}. Acknowledging the possible existence of this behaviour, all sampled reference magnetic states used in the solar dynamo correspond to the start of each Schwabe cycle, independently of the shape of the polar field curves. Hence, the true representativeness of the sampled estimated polar field strength should be assessed by the solar dynamo-NARX combined efficacy in reproducing solar activity. 

A possible way to improve the presented NARX model is by incorporating information from other recorded solar activity proxies, e.g. the record of polar faculae, to provide a broader and richer set of scenarios during the development of the model. However, in this investigation the solar cycles simulation presented in the following subsection \ref{sec:scsf}, is carried out with the models developed in this subsection using the hemispheric polar magnetic field time series alone.

\begin{figure}
 \begin{center}
\begin{tikzpicture}[spy using outlines={circle, magnification=6, connect spies}, thick,scale=1.0]
\pgfplotsset{width=8.5cm, height=6.5cm}
\begin{axis}[x tick label style={/pgf/number format/1000 sep=},
    no markers,grid=major, 
    xmin=2000, xmax=2020,
    ymin=0, ymax=250,
    xtick={1990,1995,2000,2005,2010,2015,2020},
    axis lines = left,
    xlabel = {Date [ Year ] },
    ylabel = {SSN},
    ymajorgrids=true,
    extra y tick style={grid=major, grid style={dashed,black}}, 
    major grid style={black!20},
    legend style={legend pos=north east, font=\small, legend cell align=left}
]
\addplot[color=mygreen,line width=2pt,loosely dotted] table[x index=0,y index=1,col sep=comma,mark=none] {d_ssn_obs.dat};
\addlegendentry{Observed}
\addplot[color=mylilas,line width=2pt,solid] table[x index=0,y index=1,col sep=comma,mark=none] {d_ssn_smo.dat};
\addlegendentry{Observed Smooth}
\addplot[color=red,line width=1pt,solid] table[x index=0,y index=1,col sep=comma,mark=none] {d_ssn_forec.dat};
\addlegendentry{Forecast}
\addplot[color=red,line width=0.5pt,dotted] table[x index=0,y index=2,col sep=comma,mark=none] {d_ssn_forec.dat};
\addlegendentry{3 SD Bound}
\addplot[color=red,line width=0.5pt,dotted] table[x index=0,y index=3,col sep=comma,mark=none] {d_ssn_forec.dat};
\addplot[color=mycolor5,line width=2pt,sharp plot,update limits=false,dashed] 
	coordinates {(1996.6666,0) (1996.6666,300)};
\node[text width=0.3cm, fill=lightgray!70, rounded corners] at (axis cs:1992,30) {22};
\node[text width=0.3cm, fill=lightgray!70, rounded corners] at (axis cs:2002,30) {23};	
\node[text width=0.3cm, fill=lightgray!70, rounded corners] at (axis cs:2014,30) {24};		
\coordinate (spypoint) at (axis cs:2019.4,13);
\coordinate (magnifyglass) at (axis cs:2008,130);
\end{axis}
\spy [blue, size=2.0cm] on (spypoint)
   in node[fill=white] at (magnifyglass);
\end{tikzpicture}
\caption{Start of the solar cycle 25 prediction using an autoregressive model of degree 9 on the monthly smooth SSN.}
\label{fig:ssndetail}
\end{center}
\end{figure}
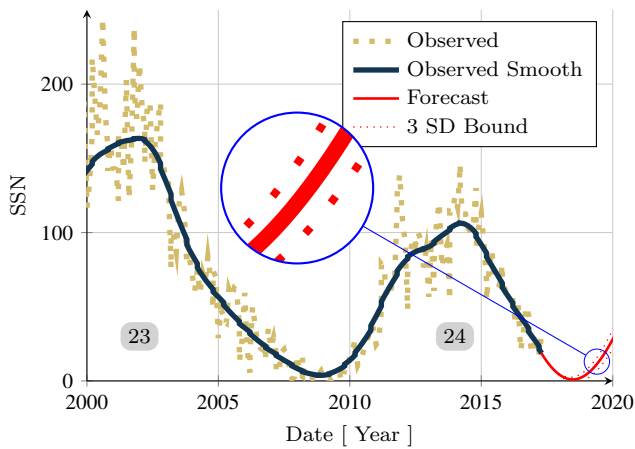

\subsection{Solar Cycles Simulation and Forecast} \label{sec:scsf}
\begin{figure*}
  \begin{center}
      \includegraphics[width=17.1cm]{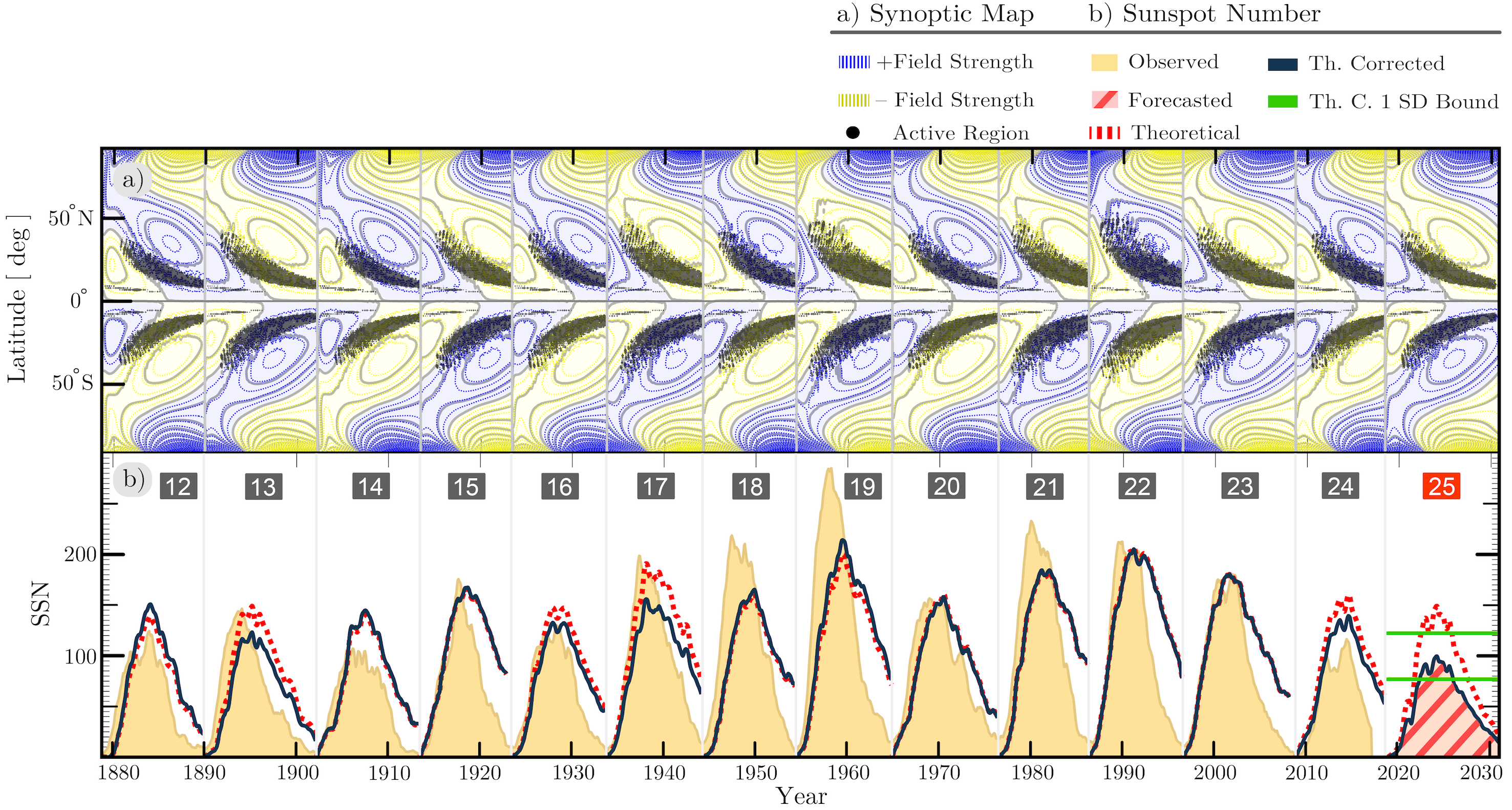}
    \caption{Solar activity simulation results. The theoretical magnetic and sunspot butterfly diagrams are shown in panel (a). Panel (b) shows the raw (dashed curves) and corrected pseudo-SSN (solid curves), superimposed on the observed 13-month smoothed monthly SSN\textsuperscript{\ref{SSNfootnote}} (light areas). Of interest in this investigation are sunspot cycles \textit{activity trends} between the simulated and the precursory cycle.}
    \label{fig:sunsynoptic}
  \end{center}
\end{figure*}

Solar activity dynamo simulations of the cycles $12-25$ are presented in Figure \ref{fig:sunsynoptic}. The theoretical magnetic and sunspot butterfly diagrams are shown in Figure \ref{fig:sunsynoptic}a, and the pseudo-SSN (solid curves) are superimposed on the observed 13-month smoothed monthly SSN (light areas) in Figure \ref{fig:sunsynoptic}b. Pseudo-SSN are obtained from the normalised and scaled chronological sum of emerging active regions in Figure \ref{fig:sunsynoptic}a, with respect to the reference observed  solar cycle 23. Note that during the declining phase, the theoretical curves overestimate the observed SSN. This is an expected behaviour as the solar dynamo model gradually adjusts itself from the imposed initial poloidal potential magnitudes and asymmetries to bring back a subsequent average solar cycle. The theoretical curves, dashed lines in Figure \ref{fig:sunsynoptic}b, show the direct output (raw) from the solar dynamo model, whereas the solid curves show the corrected theoretical curves with the process detailed below.

\begin{figure}
 \begin{center}
\begin{tikzpicture}
\pgfplotsset{width=8.5cm, height=4.9cm}
\begin{axis}[
    xmajorgrids=true,
    ymajorgrids=true,
    yminorgrids=true,
    grid=major, 
    xmin=12.5, xmax=24,
    ymin=-70, ymax=70,
    axis lines = left,
    xlabel = {Cycle Number},
    ylabel = {SSN Percentage Change},
    x axis line style= { draw opacity=0 },
    extra y tick style={grid=major, grid style={dashed,black}}, 
    major grid style={black!20},
    legend style = {at={(1.01,1.01)}, anchor = south east, legend columns = -1, draw=none, font=\small}
]
\draw [->] (axis cs:0,0) -- (axis cs:25,0) node [below] {$$};
\addplot[color=mylilas,line width=1.5pt,mark=*] coordinates {
(13,7.4144)
(14,-5.7247)
(15,18.2417)
(16,-11.0646)
(17,29.3308)
(18,-15.5094)
(19,24.2661)
(20,-19.7548)
(21,13.7445)
(22,13.2928)
(23,-13.0176)
(24,-11.1583)
};\addlegendentry{Theoretical}     
\addplot[color=mygreen,line width=1.5pt,mark=diamond*] coordinates {
(2,33.9348)
(3,36.9430)
(4,-10.9724)
(5,-65.1509)
(6,-0.9756)
(7,46.7980)
(8,105.4530)
(9,-10.2082)
(10,-15.3251)
(11,25.6713)
(12,-46.8376)
(13,17.7653)
(14,-26.8942)
(15,64.0523)
(16,-25.8964)
(17,52.5346)
(18,10.1208)
(19,30.3155)
(20,-45.0526)
(21,48.7229)
(22,-8.7591)
(23,-15.1529)
(24,-35.4409)
};\addlegendentry{Observed}   
\addplot[red,line width=1.5pt,dashed] coordinates {
    (30, -4.344602599505967)
}; \addlegendentry{Discordant}  
\draw[red,line width=2pt,->] (axis cs:22,13.2928) -- (axis cs:22,-8.7591);
\draw[red,line width=2pt,->] (axis cs:18,-15.5094) -- (axis cs:18,10.1208);
\draw[red,line width=1.5pt,dashed] (axis cs:18,-15.5094) circle (0.3cm);
\draw[red,line width=1.5pt,dashed] (axis cs:22,13.2928) circle (0.3cm);	
\end{axis}
\end{tikzpicture}
\caption{Solar maxima forward change percentage trends from the raw theoretical and observed SSN. Discordant pairs for a given solar cycle have opposite signs, whereas concordant pairs exhibit different levels of agreement. The highest level of agreement (overlapping pair) is observed in the solar cycle 23 because the theoretical response is calibrated on this cycle.}
\label{fig:suncyccomp}
\end{center}
\end{figure}
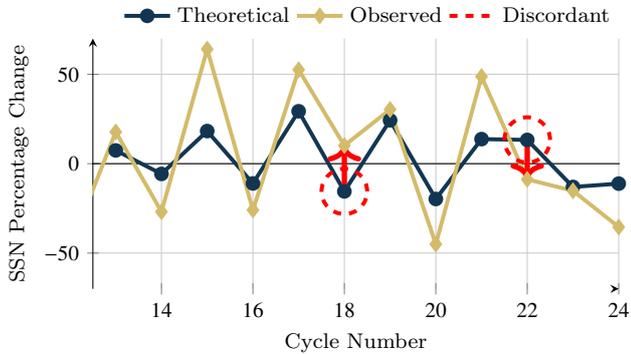

Raw data in Figure \ref{fig:sunsynoptic}b shows important characteristics of the dynamo model and its potential use for forecast application. Firstly, theoretical and observed SSN maxima of cycle 23 must match since this cycle serves as data fulcrum to relate theory with observations as discussed earlier in this section \ref{sec:solarcyfore} (refer to Figures~\ref{fig:suncalib2}-\ref{fig:suncalib1}). Other theoretical and observed maxima show close match such as in cycles 12, 17, 20, and 22; the rest show various levels of differences with the highest found in cycle 19. From these results it is evident that the solar dynamo model produces results with limited sensitivity to the simulation initial state, i.e. approximately $\pm 50$ SSN with respect to the initial observed reference solar cycle. For example, dynamo input data (solar polar field) from the observed solar cycle 20 produced underestimated pseudo-SSN activity (theoretical curve in the figure) of the cycle 21, albeit with the correct increasing trend. Due to this model limitation simulation results are assessed in pairs, that is in trend transitions from the initial reference observed cycle to the subsequent simulated cycle, rather than direct SSN cycles comparison. Figure \ref{fig:suncyccomp} summarises maxima forward percentage trends between the theoretical and observed SSN. From the theoretical points in Figure \ref{fig:suncyccomp} is seen that the transitions 17$\,\to\,$18, and 21$\,\to\,$22 are discordant to the observed trends. Or in other words, if forecast had been the case, simulation of cycles 18 and 22 would have lead to incorrect predicted trends with respect to cycles 17 and 21 respectively. 

External investigations of solar cycle activity reproduction using high diffusivity mean field axisymmetric dynamo models, show similar performances to those presented in this investigation. In a first example, in Karak's \citep{karak2010importance} study the meridional circulation is adjusted to match observed SSN periods. This parameter variation modified the average dynamo model response producing various levels of solar activity. That approach loosely approximated the magnitude of some solar cycles, with maximum difference found in the solar cycle 19. Regarding the trends between solar cycles and their precursors, cycles 15, 18, and 20 are discordant to observations in Karak's study. In another example, \citet{jiang2007solar} report simulations using the observed photospheric magnetic field as cycles' adjusting parameter, thereby encompassing cycle 21 onwards. In \citet{jiang2007solar} study, theoretical curves of the cycles 21, 22, 23 show close match to the observed SSN. Additionally a forecast for the cycle 24 is reported that is now known to have successfully anticipated peak observed activity. However, regarding the trends between solar cycles and their precursors, cycle 21 is discordant in that study. At this point is important to bear in mind that the photospheric magnetic field precursor approach by \citet{jiang2007solar}, is used as reference study in this investigation as mentioned in section \ref{sec:intro}. Similarities in the approaches of both investigations allow direct comparison of the raw results for the cycles $21-24$ in Figure \ref{fig:sunsynoptic}b. Interestingly, the discordant cycle 21$\,\to\,$22 trend and correct reproduction of the solar cycle 24 are observed in both cases. The ability to compare further solar cycles in this investigation, is enabled by the proposed polar field reconstruction of past solar cycles. The validity of the reconstructed polar field is tested in their capability to furnish adequate initial states for the solar dynamo to provide a suitable representation of cycle features. Nonetheless, the solar dynamo model as well as the reconstructed polar field add undifferentiated uncertainty components to the final solar activity simulation. Thus, theoretical cycles in this investigation report combined limitations and strengths of the solar dynamo model, the reconstructed polar field, and simulation assumptions and methodology. Comparison amongst observed and theoretical cycles in the range $12-20$ in Figure \ref{fig:sunsynoptic}b, highlights the potential of the proposed approach.

The \citet{karak2010importance} study provides an additional example of solar dynamo model performance. The range of interest for comparison in this case are cycles $12-20$. Comparison of results between Karak's study and the raw curves in Figure \ref{fig:sunsynoptic}b show that reconstructed polar field yield better solar cycles reproduction than those generated using observed SSN periods as the adjusting dynamo parameter. Karak's study shows that meridional flow variation is a fundamental ingredient in solar activity simulation. This is a reasonable argument considering that the main relationship between solar activity and meridional flow may be explained by the interaction of equatorward meridional flow at the base of the convection zone. In that possible explanation, plasma flow may encounter motion restriction imposed by magnetic tension found in highly active zones; meridional flow is quenched by solar activity. This suggests that a more robust solar dynamo model should include meridional circulation fluctuations along the solar cycles simulation. In this regard, a deeper understanding about the meridional flow structure would provide crucial information to improve current solar dynamo models. 


Reproduction of past solar cycles is fundamental to the solar dynamo validation process, which in turn improves its reliability for adequate forecasts. The analysis presented above shows that high diffusivity mean field axisymmetric dynamo models do not capture completely the whole range of observed solar activity, at least with the methodologies discussed above. A plausible explanation to the theoretical mismatch found in this investigation, is that the initial adjustment of the dynamo model includes information from only one component of the magnetic field, i.e. poloidal modulation via observed solar polar magnetic field. In this regard, simulations point to the possible need to include additional information about the state of the toroidal component. Expanding on this possibility, take the theoretical solar cycle 19 in Figure \ref{fig:sunsynoptic}b as an example. The initial adjustment of the poloidal potential in the dynamo model introduces partial information of the previous cycle 18 because, without modification, the toroidal field strength corresponds to that of an average solar cycle. With the progress of the simulated cycle, diffusion and the meridional flow gradually transports the modified poloidal field to the tachocline region where it is stretched in the azimuthal direction by the differential rotation. The preexisting and newly produced toroidal field generate emerging active regions resulting in a small increase of simulated solar activity, i.e. the solar cycle 19. If on the other hand, both fields are adjusted, the newly created toroidal field would find a suitable magnitude of the preexisting toroidal field to effectively produce the observed activity. The challenge in this case, however, is the identification of an observed proxy of the toroidal field state. 

\begin{figure} 
 \begin{center}
\begin{tikzpicture}
\pgfplotsset{width=8.5cm, height=4.9cm}
\begin{axis}[
    xmajorgrids=true,
    ymajorgrids=true,
    yminorgrids=true,
    grid=major, 
    xmin=12.5, xmax=24,
    ymin=-70, ymax=70,
    axis lines = left,
    xlabel = {Cycle Number},
    ylabel = {SSN Percentage Change},
    x axis line style= { draw opacity=0 },
    extra y tick style={grid=major, grid style={dashed,black}}, 
    major grid style={black!20},
    legend style = {at={(1.01,1.01)}, anchor = south east, legend columns = -1, draw=none, font=\small}
]
\draw [->] (axis cs:0,0) -- (axis cs:25,0) node [below] {$$};
\addplot[color=mylilas,line width=1.5pt,mark=*] coordinates {
(13,-18.2346)
(14,17.5304)
(15,15.5404)
(16,-20.7355)
(17,17.4806)
(18,6.0687)
(19,29.5739)
(20,-26.2329)
(21,16.7407)
(22,11.2268)
(23,-12.1354)
(24,-22.7237)
};\addlegendentry{Theoretical}     
\addplot[color=mygreen,line width=1.5pt,mark=diamond*] coordinates {
(2,33.9348)
(3,36.9430)
(4,-10.9724)
(5,-65.1509)
(6,-0.9756)
(7,46.7980)
(8,105.4530)
(9,-10.2082)
(10,-15.3251)
(11,25.6713)
(12,-46.8376)
(13,17.7653)
(14,-26.8942)
(15,64.0523)
(16,-25.8964)
(17,52.5346)
(18,10.1208)
(19,30.3155)
(20,-45.0526)
(21,48.7229)
(22,-8.7591)
(23,-15.1529)
(24,-35.4409)
};\addlegendentry{Observed}    
\addplot[red,line width=1.5pt,dashed] coordinates {
    (30, -4.344602599505967)
}; \addlegendentry{Discordant} 
\draw[red,line width=2pt,->] (axis cs:22,11.2268) -- (axis cs:22,-8.7591);
\draw[red,line width=1.5pt,dashed] (axis cs:22,11.2268) circle (0.3cm);
\draw[red,line width=2pt,->] (axis cs:14,17.5304) -- (axis cs:14,-26.8942);
\draw[red,line width=1.5pt,dashed] (axis cs:14,17.5304) circle (0.3cm);
\draw[red,line width=2pt,->] (axis cs:13,-18.2346) -- (axis cs:13,17.7653);
\draw[red,line width=1.5pt,dashed] (axis cs:13,-18.2346) circle (0.3cm);	
\end{axis}
\end{tikzpicture}
\caption{Same as Figure \ref{fig:suncyccomp} with corrected theoretical SSN. Each cycle correction factor improves the level of agreement observed in the cycle pairs. In addition, possible non-uniform or faulty formatted data in the early cycles record is disclosed.}
\label{fig:suncyccomp2}
\end{center}
\end{figure}
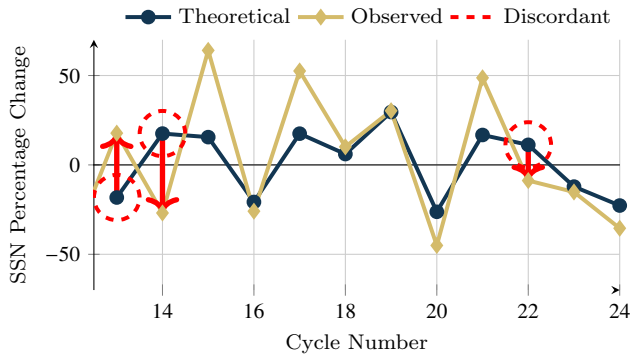

The rise times of solar cycles may give qualitative information of the toroidal field state. This hypothesis is based on an extreme case wherein the preexisting saturated toroidal field at the beginning of the solar cycle, triggers large flux tube buoyant sections that rise rapidly to the solar surface. In contrast to the poloidal field, this toroidal field variation is slow, possibly sufficient to retain its bulk characteristics even after maximum solar activity. If this is true, then a given solar cycle rise time and the polar magnetic field at the end of that cycle, may contain sufficient information to adjust the solar dynamo model for the simulation of the subsequent solar cycle. In order to capture the qualitative state of the toroidal field into a quantitative correction factor, Eq. \ref{toroadjust} is used wherein $t_i$, $t_f$, and $t_{max}$ are solar cycles times at the start, end, and activity peak respectively.
\begin{equation}\label{toroadjust}
C = \frac{t_f - t_{max}}{t_f-t_i}
\end{equation}
The correction is carried out in cycle pairs wherein the cycle $(n -1)$ provides the factor $C$ for the cycle $n$. For example, to preserve the magnitude of the reference solar cycle 23, the correction vector is normalised to the cycle 22 as this sets the initial conditions of the cycle 23. The normalised correction vector is then multiplied by the raw theoretical curves shown in Figure \ref{fig:sunsynoptic}b. The corrected theoretical cycles are presented in the same figure as solid curves. This modification results in a dramatic improvement on the magnitudes and trends of simulated cycles with respect to observed activity. Figure \ref{fig:suncyccomp2}, shows the corrected trends of the simulated cycles. SSN percentage change error in solar cycle 22 is reduced in comparison to the same cycle in Figure \ref{fig:suncyccomp}, and the discordant trend in cycle 18 disappears. Nonetheless, two new discordant trends appear in cycles 13 and 14. Interestingly, the erroneous cycle transitions, i.e. 12$\,\to\,$13 and 13$\,\to\,$14, coincide with the Maunder's original butterfly diagrams published in 1904 covering the period $1876-1902$ \citep{maunder1904note}. In Maunder's original work, butterfly diagrams were built with data of sunspots permanence at a given latitude. Wherever a sunspot centre appeared on the solar surface, and remained for more than one day during a given solar rotation, the latitude was marked in a latitude-time chart. Sunspot area information was added later with the assembly of sunspot observations as formats changed along the years. Recent evaluation of early datasets has identified non-uniform sunspot area data\textsuperscript{\ref{BFDfootnote}}. In an effort to regularise datasets, these were compared with the Mount Wilson photographic plate collection dating from 1917 to 1941\textsuperscript{\ref{BFDfootnote}}. All this may suggest that the anomalous transition trends between theoretical and observed data for cycles 13 and 14 in Figure \ref{fig:suncyccomp2}, may be a result of non-uniform or erroneously formatted sunspot area data.

The solar dynamo model's adequate reproduction of solar peak activity and solar maxima forward change percentage trends of past solar cycles, allows a forecast for the solar cycle 25 to be proposed. A weak solar cycle 25, with slow rise time and maximum activity $-14.4\%$ ($\pm 19.5\%$) with respect to the current cycle 24 is predicted.

\section{Conclusions}
In this paper we have analysed the capability of a mean field axisymmetric dynamo model with a manipulated magnetic field to reproduce observed solar activity. In general, current solar activity reproduction using solar dynamo models with a poloidal field precursor is limited to a few solar cycles. Aiming at extending the verification of the solar dynamo model with a broader number of solar activity scenarios, unrecorded solar polar magnetic field were reconstructed. Reconstruction of the solar polar magnetic field was carried out using the hemispheric sunspot areas record in conjunction with artificial neural network methods. As a result, the extended solar dynamo model testing encompassed thirteen solar cycles, rather than the conventional observed four cycles. 

Simulations of 13 solar cycles with the proposed diffusion-dominated solar dynamo approach, showed satisfactory reproduction of SSN peak trends for 9 of 12 solar cycles. In addition, the extended verification permitted the identification of a correction factor to improve solar dynamo simulations. Further, possible anomalous data in the historical hemispheric sunspot area are identified in early solar cycle reports. Corrected simulated activity for the forthcoming solar cycle 25 predicts weaker peak activity than observed in the current cycle.

Solar activity simulations presented in this investigation should be treated as rough approximations as many poorly understood physical processes controlling the actual solar cycle are represented in the model. Nonetheless, the reported combination of the consolidated solar polar magnetic field strengths, mean field axisymmetric dynamo model parameterisation approach, and calibration methodology, proved to be sufficiently effective in reproducing various historical solar activity scenarios.

\bibliographystyle{mnras}
\bibliography{references_asph}

\appendix
\section{Supplementary material}
\label{sec:suppmat}

\begin{figure}
      \includegraphics[width=8.4cm]{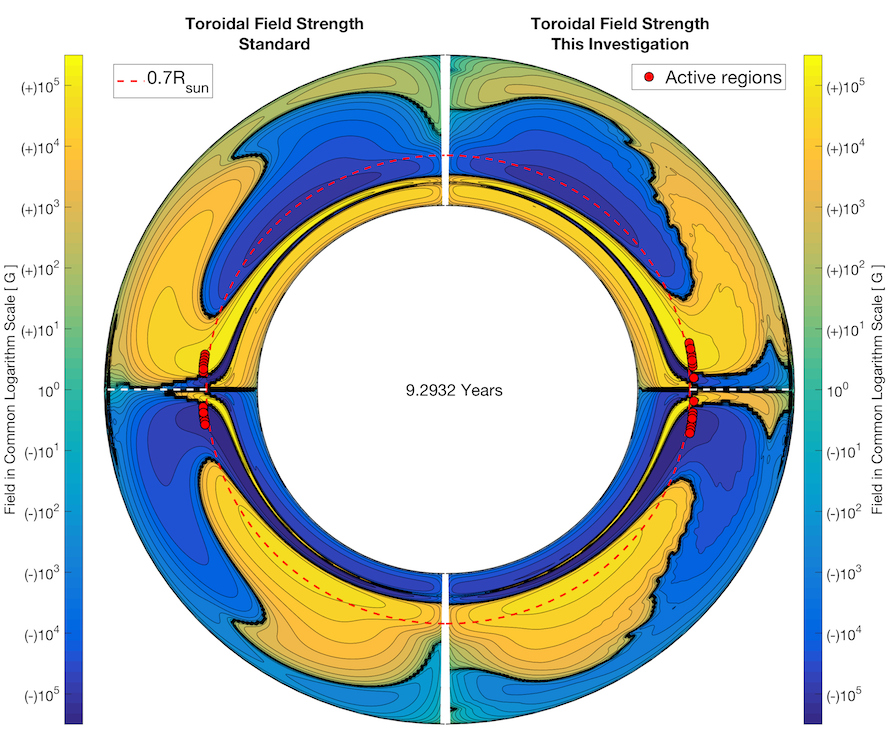}
    \caption{A snapshot of the supplementary video. Comparison of simulated toroidal field morphology evolution using standard and the proposed solar parameterisations.}
    \label{fig:suppmaterial}
\end{figure}

\bsp	
\label{lastpage}
\end{document}